\documentclass{article}
\usepackage[dvips]{graphicx}
\usepackage{amsmath}
\usepackage{amsfonts}
\usepackage{amssymb}

\begin{document}

\title{Designing agent-based market models}
\author{Paul Jefferies and Neil F. Johnson\\Physics Department, Oxford University\\Clarendon Laboratory, Parks Road, Oxford OX1 3PU, U.K.\\\emph{p.jefferies@physics.ox.ac.uk \ \ \ n.johnson@physics.ox.ac.uk}}
\maketitle
\begin{abstract}
In light of the growing interest in agent-based market models, we bring
together several earlier works in which we considered the topic of
self-consistent market modelling. Building upon the binary game structure of
Challet and Zhang, we discuss generalizations of the strategy reward scheme
such that the agents seek to maximize their wealth in a more direct way. We
then examine a disturbing feature whereby such reward schemes, while appearing
microscopically acceptable, lead to unrealistic market dynamics (e.g.
instabilities). Finally, we discuss various mechanisms which are responsible
for re-stabilizing the market in reality. This discussion leads to a `toolbox'
of processes from which, we believe, successful market models can be
constructed in the future.
\end{abstract}

\vskip0.5in

\noindent Oxford Center for Computational Finance working paper: \emph{OCCF/010702}

\newpage

\section{Introduction\label{sec intro}}

Agent-based models have attracted significant interest across a broad range of
disciplines \cite{1arthur}. An increasingly popular application of these
models has been the study of financial markets \cite{1arthur,2econophysics}.
The motivations for this focus on financial markets are two-fold: the promise
that data-rich financial markets are good candidates for the empirical study
of complex systems, and the inadequacy of standard economic models based on
the notions of equilibria and rational expectations. Currently many different
agent-based models exist in the Econophysics literature, each with its own set
of implicit assumptions and interesting properties
\cite{1arthur,2econophysics,3challet}. In general these models manage to
exhibit some of the statistical properties that are reminiscent of those
observed in real-world financial markets; for example, fat-tailed
distributions of returns and long-timescale volatility correlations. \ Despite
their differences, these models draw on several of the same key ideas:
feedback, frustration, adaptability and evolution. The underlying goal of all
this research effort is to generate a microscopic agent-based model which (i)
reproduces all the stylized facts of a financial market, \emph{and} (ii) makes
sense on the microscopic level in terms of financial market microstructure.
While each of these goals is separately achievable, the combination of (i)
\emph{and} (ii) within a single model represents a fascinating challenge, and
underpins our motivation for writing the present paper.

Challet and Zhang's Minority Game \cite{3challet} offers possibly the simplest
paradigm for a system containing the key features of feedback, frustration and
adaptability. The MG has remarkably rich dynamics, given the simplicity of its
underlying binary structure, and has potential applications in a wide range of
fields. Consequently Challet and Zhang's MG model, together with the original
bar model of Arthur \cite{1arthur}, represent a major milestone in the study
of complex systems. Given the MG's richness and yet underlying simplicity, the
MG has also received much attention as a financial market model
\cite{2econophysics}. The MG comprises an odd number of agents \emph{N}
choosing repeatedly between the options of buying (1) and selling (0) a
quantity of a risky asset. The agents continually strive to make the minority
decision i.e. buy assets when everyone else is selling and sell when everyone
else is buying. This strategy seems to make sense on first inspection. For
example, a majority of agents selling will force the price of the asset down
and thus ensure a low buy price (which is favorable) for the minority of the
agents who decided to buy; hence the minority group has `won'. However, now
consider the scenario where this pattern repeats over and over. The minority
of agents, who have been buying more and more assets as the price falls, now
find themselves holding a huge inventory of assets which are worth very
little. When these agents try to cash in their assets, they will find
themselves very much poorer. \emph{Surely then the minority group has actually lost?}

This paradox has sparked much discussion as to the suitability of the MG as
the basis of a microscopic model of financial markets. Consequently a diverse
range of modified (yet MG-related) models have emerged, all seeking to address
this point in different ways \cite{2econophysics}. Some models include a
combination of different types of agents - in particular, agents who are
rewarded for trading in the minority, and agents who are rewarded oppositely
for trading in the majority (see Ref. \cite{4liege} for an early example). In
such models, the overall class of the game (minority or majority) is
ambiguous, and is likely to change as differing numbers of each group flood in
and out of the market. Other models try to combine the two different reward
structures, thus requiring only one class of agents who strive to be neither
exclusively in the minority group nor exclusively in the majority group
\cite{5dollar,6bouchaudlong,7occf}. We refer the reader to the Econophysics
website \cite{2econophysics} for further examples (see also Refs.
\cite{challet,15marsili,farmer,8dublin}).

The purpose of this paper is to consolidate the search for a suitable
microscopic reward scheme for use in agent-based financial markets. The paper
comprises a synthesis of several earlier reports, each aimed at addressing the
topic of self-consistent agent-based market modelling within a binary game
structure. These earlier reports are re-organized in light of several recent
papers on specific reward structures, and supplemented by additional
simulation results, observations and thoughts \cite{7occf}. We start by
discussing various strategy reward schemes in which the agents seek to
maximize their wealth in a direct way. We then illustrate a disturbing feature
whereby such reward structures, while appearing microscopically acceptable,
can lead to unrealistic market dynamics (e.g. instabilities). Finally, we
discuss various mechanisms which are responsible for re-stabilizing the market
in reality. This discussion leads to a `toolbox' of processes from which, we
believe, successful market models can be constructed in the future.

\section{Binary game structure\label{sec MG basics}}

Consider the situation of a population of $N$ agents with some limited global
resource. A specific example could be Arthur's famous bar problem
\cite{1arthur} where there are $N$ regular bar-goers but only $L<N$ seats. The
reward scheme in Arthur's bar problem is simple: bar-goers are successful if
they attend \emph{and} they manage to obtain a seat. This reward scheme
implicitly assumes that the bar-goers value sitting down above other criteria.
While this makes sense for many situations, it is not universal. For example,
it is our experience that customers of college-based bars do not view seating
as a necessary requirement for a successful evening! The motto `the more the
merrier' often seems more appropriate. The MG represents a very special case
of the bar problem, in which the seating capacity $L=(N-1)/2$ \emph{and} all
the attendees wish to sit down. In more general and realistic situations, the
correct reward scheme is likely to be less simple than either Arthur's bar
model or Challet and Zhang's MG. Until a specific reward scheme is defined,
the bar model remains ill-specified. Putting this another way, the precise
reward scheme chosen is a fundamental property of the resulting system and
directly determines the resulting dynamical evolution. Hence it is crucial to
understand the effect that different microscopic reward schemes have on the
macroscopic market dynamics.

As first discussed in Ref. \cite{1arthur}, the bar problem is somewhat
analogous to a financial market where the bar-goers are replaced by traders.
In the same way that a general bar problem requires a non-trivial reward
scheme, any `correct' agent-based market model will need a non-trivial reward
scheme in order to avoid inconsistencies with financial market microstructure.
This provides us with the motivation to look beyond the bar model and MG, in
order to design an agent-based model with financial relevance. Having said
this, the binary structure of the MG provides an appealing framework for
formulating a market model without introducing too many obvious pathologies.
Given these considerations, our discussion from here on will not assume any
specific reward scheme, but will employ a binary game structure based on Ref.
\cite{3challet}.

We start by reviewing the binary game structure. There are three basic
parameters; the number of agents $N$, the `memory' of the agents $m$, the
number of strategies held by each agent $s$. The agents all observe a common
binary source of information, but only remember the previous $m$ bits. Hence
the global information available to each agent at time $t$ is given by
$\mu\left[  t\right]  $ where in decimal notation $\mu\left[  t\right]
\in\left\{  0\ldots P-1\right\}  $ with $P=2^{m}$. Each strategy
$\underline{a}_{R}$ contains as its elements $a_{R}^{\mu}$, and represents a
response $\left\{  -1,1\right\}  $ to each of the possible $P$ values of the
global information $\mu$. There are hence $2^{P}$ possible strategies. The
agents are randomly allocated a subset of $s$ of these strategies at the
outset of the game, and are not allowed to replace these. The agents also keep
a score $S_{R}$ reflecting their strategies' previous successes, whether the
strategy is played or not. The net action of the agents $A\left[  t\right]  $
is defined as:
\[
A\left[  t\right]  =\sum_{R=1}^{2^{P}}n_{R}\left[  t\right]  a_{R}^{\mu\left[
t\right]  }%
\]
where $n_{R}$ is the number of agents playing strategy $R$. The agents always
play their highest performing strategy, i.e. the strategy with the highest
$S_{R}$. The strategy score is updated with payoff $g_{R}$ such that
$S_{R}\left[  t+1\right]  =S_{R}\left[  t\right]  +g_{R}\left[  t+1\right]  $.

The central question is the following: what should we take as `success' in the
context of a market and hence what are the payoffs? In short, what is the
microscopic reward scheme? This is equivalent to asking what the game actually
is, since a game requires a definition of the players' goals and hence their
rewards. In order to address this question, we start by considering a binary
version of Arthur's bar problem with general $L$, i.e. an MG-like model which
is generalized to $L\neq(N-1)/2$ \cite{usvarL1,usvarL2}. The two responses
would be `attend' or `don't attend', and there are two possible scenarios: (i)
the attendance is less than the seating capacity. In this case, the attendees
are successful while those who didn't attend are unsuccessful. (ii) The
attendance exceeds the seating capacity. In this case, the attendees are
unsuccessful while those who didn't attend are successful. Setting $L=(N-1)/2$
yields the MG, and the two scenarios collapse into one: the successful agents
are those who predict the minority group, i.e. attend when the minority
attend, or don't attend when the majority attend. [For a study of the binary
game with $L\neq(N-1)/2$, see Ref. \cite{usvarL1,usvarL2}.] In the specific
case of the MG, success is classified as having correctly predicted the
minority outcome. The payoff function for the MG is thus of the form:
\begin{equation}
g_{R}\left[  t+1\right]  =\chi\left[  -a_{R}^{\mu\left[  t\right]  }A\left[
t\right]  \right]  \label{eq strat payoff MG}%
\end{equation}
where $\chi\left[  x\right]  $ is an odd, increasing function of $x$ usually
chosen to be either $\chi\left[  x\right]  =x$ or $\chi\left[  x\right]
=\operatorname*{sgn}\left[  x\right]  $. The feedback in the MG arises through
the global information $\mu\left[  t\right]  $ where the most recent bit of
the binary information is defined by the sign of the net action of the agents
$A\left[  t\right]  $:
\begin{equation}
\mu\left[  t+1\right]  =2\mu\left[  t\right]  -P\operatorname{H}\left[
\mu\left[  t\right]  -\frac{P}{2}\right]  +\operatorname{H}\left[  A\left[
t\right]  \right]  \label{eq history update MG}%
\end{equation}
where $\operatorname*{H}\left[  x\right]  $ is the Heaviside function
\cite{uspre}. In the more general form of the game where the resource level
$L\neq(N-1)/2$, the argument in the Heaviside function $\operatorname{H}%
\left[  \dots\right]  $ would depend on the value of $L$ \cite{usvarL2}. More
generally it could become a complicated function of all the game parameters,
or include feedback from the macroscopic dynamics in the past, or even the
effects of exogenous news arrival.

It is now accepted that a generalization of the MG in which agents can sit out
of the game (i.e. $a^{\mu\left[  t\right]  }=0$) when they are not confident
of the success of any of their $s$ strategies, provides a better model of
financial trading since it allows the number of agents active in the market to
vary. This more general model is the Grand Canonical Minority Game (GCMG)
which was first studied in Ref. \cite{8dublin}. The GCMG typically has two
more parameters than the MG: namely $T$ which is a time horizon over which
strategy scores are forgotten (i.e. $S_{R}\left[  t+1\right]  =\left(
1-1/T\right)  S_{R}\left[  t\right]  +g_{R}\left[  t+1\right]  $) and $r$
which is a threshold value of $S_{R}$ below which the agents will opt not to
participate in the game ($a^{\mu\left[  t\right]  }=0$). The GCMG seems to
reproduce the stylized statistical features of a financial time-series over a
wide parameter range (see the discussion in Ref. \cite{9bouchaudprague}).

\section{The price formation process\label{sec price process}}

It is commonly believed that in a financial market, excess demand (i.e. the
difference between the number of assets offered and the number sought by the
agents) exerts a force on the price of the asset. Furthermore it is believed
that a positive excess demand will force the price up and a negative demand
will force the price down. A reasonable suggestion for the price formation
process could then be \cite{10contbf}:
\begin{equation}
\ln\left[  p\left[  t\right]  \right]  -\ln\left[  p\left[  t-1\right]
\right]  =\frac{D\left[  t^{-}\right]  }{\lambda}%
\label{eq price formation mult}%
\end{equation}

or
\begin{equation}
p\left[  t\right]  -p\left[  t-1\right]  =\frac{D\left[  t^{-}\right]
}{\lambda}\label{eq price formation add}%
\end{equation}
where $D\left[  t^{-}\right]  $ represents the excess demand in the market
prior to time $t$, when the new price $p\left[  t\right]  $ is set and the
buy/sell orders are executed. The scale parameter $\lambda$ represents the
`market depth' (or liquidity) i.e. how sensitive a market is to an order
imbalance. In general we would expect $\lambda$ to be some increasing function
of the number of traders $N$ trading in that asset. Also, the functional forms
may not be linear as in Equations \ref{eq price formation mult} and \ref{eq
price formation add}. However, studies carried out on several markets have
found the forms of Equations \ref{eq price formation mult} and \ref{eq price
formation add} to be reasonable \cite{11empiricaldollar}.

In Ref. \cite{4liege}, we suggested that in markets involving a market-maker
who takes up the imbalance of orders between buyers and sellers, the
market-maker will wish to manipulate the price of the asset in order to
minimize her inventory of stock $\phi_{M}\left[  t\right]  $. Whilst the
market-maker's job is to execute as many of the orders as is possible, she
does not wish to accumulate a high position in either direction. To this end,
the market-maker lowers the price to attract buyers when she holds a long
position $\phi_{M}>0$ and raises it to attract profit-taking sellers when she
has a short position $\phi_{M}<0$. This modifies the price formation of
Equation \ref{eq price formation mult} to\footnote{It is important to note
that this price process is not abitrage-free i.e. it is possible for the
agents to make profit by manipulating the market maker.}:
\begin{equation}
\ln\left[  P\left[  t\right]  \right]  -\ln\left[  P\left[  t-1\right]
\right]  =\frac{D\left[  t^{-}\right]  }{\lambda}-\frac{\phi_{M}\left[
t^{-}\right]  }{\lambda_{M}}\label{eq price formation mult MM}%
\end{equation}
where $\lambda_{M}$ is the market-maker's sensitivity to her inventory.

\section{From MG to market model\label{sec MG to MM}}

Our puzzle is to form a sensible model of a collection of traders buying and
selling a financial asset. It is clear from Sections \ref{sec MG basics} and
\ref{sec price process} that we have two pieces of the puzzle. The MG gave us
a simple paradigm for a group of agents interacting in response to global
information and adapting their behavior based on past experience. The output
of this model was a `net action' $A\left[  t\right]  $. In addition, we now
have some simple systems for price-formation based on an excess demand
$D\left[  t\right]  $. But what relationship should we assume between
$D\left[  t\right]  $ and $A\left[  t\right]  $? To answer this question, we
should clarify what we want the individual actions $a_{R}^{\mu\left[
t\right]  }$ of the strategies to resemble. In short, the question is `what
does a strategy map from and to?'

In most studies of the MG as a market model, researchers have taken the
actions $a_{R}^{\mu\left[  t\right]  }\in\left\{  -1,1\right\}  $ to be the
actions of selling and buying a quanta of the asset respectively. Thus the net
action of all agents $A\left[  t\right]  $ naturally becomes the excess
demand, i.e. we have:
\begin{equation}
D\left[  t^{-}\right]  =A\left[  t-1\right]  \label{eq demand MG}%
\end{equation}
Given this and the fact that the price change is an increasing function of
excess demand (for Equations \ref{eq price formation mult} and \ref{eq price
formation add}), the global information becomes the directions (up or down) of
the previous price changes. So the answer here is that a strategy maps from
the history of past asset price movements to a buy/sell signal. A strategy
mapping \emph{from} the history of past asset price movements seems a natural
and sensible choice, since financial chartists do look at precisely this
information to decide upon a course of action. We therefore propose that in
general, for this class of market model, we should always use the following
Equation (Eq. \ref{eq history update Gen}) to update the global information,
irrespective of whether Equation \ref{eq demand MG} holds.
\begin{equation}
\mu\left[  t+1\right]  =2\mu\left[  t\right]  -P\operatorname{H}\left[
\mu\left[  t\right]  -\frac{P}{2}\right]  +\operatorname{H}\left[  \Delta
P\left[  t+1,t\right]  \right]  \label{eq history update Gen}%
\end{equation}

It is worth now spending a few moments considering the timings in this model.
Equation \ref{eq demand MG} is for the excess demand at time $t^{-}$, i.e.
prior to time $t$ when all the orders are processed and a new price $p\left[
t\right]  $ is formed. Thus $D\left[  t^{-}\right]  $ can only result from all
the information that is available at time $t-1$, i.e. $\mu\left[  t-1\right]
$ and $\left\{  S_{R}\left[  t-1\right]  \right\}  $. From this information,
the agents take actions $a^{\mu\left[  t-1\right]  }$ producing a net action
of $A\left[  t-1\right]  $. The agents' actions (orders) however do not get
realized (executed) until time $t$ when the new price $p\left[  t\right]  $ is
known. The MG payoff function of Equation \ref{eq strat payoff MG}, hence
rewards agents positively for deciding to sell (buy) assets (i.e.
$a^{\mu\left[  t-1\right]  }=-1$ $\left(  1\right)  $) when the order
execution price $p\left[  t\right]  $ is above (below) the price level at the
time they made the decision (i.e. $p\left[  t-1\text{.}\right]  $).

Why is this mechanism of moving in the minority a physically reasonable
ambition for the agents? Let us consider the `notional' wealth of an agent $i
$ to be given by:
\begin{equation}
W_{i}\left[  t\right]  =\phi_{i}\left[  t\right]  p\left[  t\right]
+C_{i}\label{eq agent wealth}%
\end{equation}
where $\phi_{i}$ is the number of assets held and $C_{i}$ is the amount of
cash held. It is clear from Equation \ref{eq agent wealth} that an exchange of
cash for asset at any price does not in any way affect the agents' notional
wealth. However, the point is in the terminology. The wealth $W_{i}\left[
t\right]  $ is only \emph{notional} and not real in any sense - the only real
measure of wealth is $C_{i}$ which is the amount of capital the agent has
available to spend. Thus it is evident that an agent has to do a `round trip'
(i.e. buy (sell) an asset then sell (buy) it back) to discover whether a
\emph{real} profit has been made. Let us consider two examples of such a round
trip. In the first case the agent trades in the minority while in the second
he trades in the majority:

\begin{itemize}
\item {\small Moving in minority:}%

\begin{tabular}
[c]{|l|l|l|l|l|l|}\hline
$t$ & Action $a\left[  t\right]  $ & $C_{i}\left[  t\right]  $ & $\phi
_{i}\left[  t\right]  $ & $p\left[  t\right]  $ & $W_{i}\left[  t\right]
$\\\hline
1 & Submit buy order & 100 & 0 & 10 & 100\\\hline
2 & Buy..., Submit sell order & 91 & 1 & 9 & 100\\\hline
3 & Sell & 101 & 0 & 10 & 101\\\hline
\end{tabular}

\item {\small Moving in majority:}%

\begin{tabular}
[c]{|l|l|l|l|l|l|}\hline
$t$ & Action $a\left[  t\right]  $ & $C_{i}\left[  t\right]  $ & $\phi
_{i}\left[  t\right]  $ & $p\left[  t\right]  $ & $W_{i}\left[  t\right]
$\\\hline
1 & Submit buy order & 100 & 0 & 10 & 100\\\hline
2 & Buy..., Submit sell order & 89 & 1 & 11 & 100\\\hline
3 & Sell & 99 & 0 & 10 & 99\\\hline
\end{tabular}
\end{itemize}

\noindent As can be seen, moving in the minority creates wealth for the agent
upon completion of the necessary round-trip, whereas moving in the majority
loses wealth. However if the agent had held the asset for any length of time
between buying it and selling it back, his wealth would also depend on the
rise and fall of the asset price over the holding period. Thus the MG
mechanism seems perfectly reasonable for a collection of traders who simply
buy/sell on one timestep and sell/buy back on the next, but this is not of
course what real financial traders do in general. This pinpoints the main
criticism of the MG as a market model.

\section{Modified payoffs\label{sec payoff mods}}

The minority mechanism at play in the MG, arises solely from Equation \ref{eq
strat payoff MG} for the payoff (reward) $g_{R}\left[  t+1\right]  $ given to
each strategy $R$ based on its action $a_{R}^{\mu\left[  t\right]  }$. Thus it
can be trivially conjectured that changing the structure of the payoff
function will change the class of game being played. In Section \ref{sec MG to
MM}\ we pointed out that although \emph{trading} in the minority was
beneficial, the minority payoff structure itself made no allowance for the
rise or fall in the value of the agent's portfolio $\phi_{i}$ of assets. Let
us try to rectify this by examining the form of the agent's notional wealth,
Equation \ref{eq agent wealth}. If we differentiate the notional wealth, we
get an expression for $\Delta W_{i}\left[  t+1,t\right]  =W_{i}\left[
t+1\right]  -W_{i}\left[  t\right]  $ given by:
\[
\Delta W_{i}\left[  t+1,t\right]  =\Delta C_{i}\left[  t+1,t\right]  +p\left[
t+1\right]  \Delta\phi_{i}\left[  t+1,t\right]  +\phi_{i}\left[  t\right]
\Delta p\left[  t+1,t\right]
\]
The first two terms cancel because the amount of cash lost $-\Delta
C_{i}\left[  t+1,t\right]  $ is used to buy the extra $\Delta\phi_{i}\left[
t+1,t\right]  $ assets at price $p\left[  t+1\right]  $. This leaves us with:
\begin{equation}
\Delta W_{i}\left[  t+1,t\right]  =\phi_{i}\left[  t\right]  \Delta p\left[
t+1,t\right] \label{eq agent wealth chng}%
\end{equation}
We can then use Equation \ref{eq agent wealth chng} to work out an appropriate
reward $g_{R}\left[  t+1\right]  $ for each strategy based on whether its
action $a_{R}^{\mu\left[  t\right]  }$ induced a positive or negative increase
in notional wealth. Let us first use the fact that the price change $\Delta
p\left[  t+1,t\right]  $ is roughly proportional to the excess demand
$D\left[  \left(  t+1\right)  ^{-}\right]  $: this can be seen explicitly from
our earlier equation for the price formation, Equation \ref{eq price formation
add}. Then let's use our MG interpretation of the net action $A\left[
t\right]  $, Equation \ref{eq demand MG}. We thus have from Equation \ref{eq
agent wealth chng}:
\begin{align*}
\Delta W_{R}\left[  t+1,t\right]   &  \varpropto\phi_{R}\left[  t\right]
D\left[  \left(  t+1\right)  ^{-}\right] \\
&  \varpropto\phi_{R}\left[  t\right]  A\left[  t\right]
\end{align*}
We then identify the accumulated position of a strategy at time $t$, $\phi
_{R}\left[  t\right]  $, to be the sum of all the actions (orders) made by
that strategy which have been executed within times $0\ldots t$. Remembering
that at time $t$, the action (order) $a_{R}^{\mu\left[  t\right]  }$ has
\emph{not} yet been executed (it gets executed at $t+1$), this gives $\phi
_{R}\left[  t\right]  =\sum_{i=0}^{t-1}a_{R}^{\mu\left[  i\right]  }$. Let us
then set the payoff given to a strategy $g_{R}\left[  t+1\right]  $, to be an
increasing (odd) function $\chi$ of the notional wealth increase $\Delta
W_{R}\left[  t+1,t\right]  $ for that strategy. We thus arrive at:
\begin{equation}
g_{R}\left[  t+1\right]  =\chi\left[  \sum_{i=0}^{t-1}a_{R}^{\mu\left[
i\right]  }\times A\left[  t\right]  \right] \label{eq strat payoff
NWealth}%
\end{equation}
We could also propose a locally-weighted equivalent of Equation \ref{eq strat
payoff NWealth}, in which the reward given to a strategy is more heavily
weighted on the result of its recent actions than the actions it made further
in the past. This gives
\begin{equation}
g_{R}\left[  t+1\right]  =\chi\left[  \sum_{i=0}^{t-1}\left(  1-1/T\right)
^{t-1-i}a_{R}^{\mu\left[  i\right]  }\times A\left[  t\right]  \right]
\label{eq strat payoff NWealth T}%
\end{equation}
where $T$ enumerates a characteristic time-scale over which the position
accumulated by the strategy is `forgotten'. The $T=1$ limit of Equation
\ref{eq strat payoff NWealth T} yields $g_{R}\left[  t+1\right]  =\chi\left[
a_{R}^{\mu\left[  t-1\right]  }A\left[  t\right]  \right]  $, i.e. only the
position resulting from the most recently executed trade is taken into
account. With $T=1$, this payoff structure essentially rewards a strategy at
time $t+1$ based on whether the notional wealth change $\Delta W_{R}\left[
t+1,t\right]  $ was \emph{more positive} than it would have been if action
$a_{R}^{\mu\left[  t-1\right]  }$\ had \emph{not} been taken.

If Equation \ref{eq strat payoff NWealth T} is used in an agent-based market
model, the agents play the strategy they hold which has accumulated the
highest `virtual' notional wealth. We mean `virtual' in the sense that the
strategy itself will not have \emph{actually} accumulated this notional wealth
unless it has been played without fail since time $t=0$. The agents in this
model are thus all striving to increase their notional wealth and are allowed
to do so by taking arbitrarily large positions $\phi_{i}$. \textbf{We will
call this incarnation of the market model \$G\ref{eq strat payoff NWealth T}}:
the term `\$G' is named after the `The \$-Game' of Ref. \cite{5dollar} where
agents also strive to maximize notional wealth, and `\ref{eq strat payoff
NWealth T}' is named after Equation \ref{eq strat payoff NWealth T} for the
strategy payoff. We stress that this incarnation of the market model is by no
means unique. For example, it may be more appropriate to consider a model
wherein each agent is only allowed one position in the asset at any time, i.e.
$\phi_{i}\left[  t\right]  =-1,0,1$. This amounts to us redefining the
strategies $R$ as mapping to the actions `go long/short' instead of
`buy/sell', i.e. $a_{R}^{\mu\left[  t\right]  }=\phi_{R}\left[  \left(
t+1\right)  ^{-}\right]  $. Note the reference to $\left(  t+1\right)  ^{-}$
here because the action $a_{R}^{\mu\left[  t\right]  }$ is not executed until
time $t+1$: after this time $a_{R}^{\mu\left[  t\right]  }=\phi_{R}\left[
\left(  t+1\right)  \right]  .$ In this new model, the net action of the
agents $A\left[  t\right]  $ therefore represents the overall desired position
of the population at time $\left(  t+1\right)  ^{-}$, i.e. $A\left[  t\right]
=\sum_{i=1}^{N}\phi_{i}\left[  \left(  t+1\right)  ^{-}\right]  $. It
therefore follows that the excess demand is given by:
\begin{align}
D\left[  \left(  t+1\right)  ^{-}\right]   &  =\sum_{i=1}^{N}\left(  \phi
_{i}\left[  \left(  t+1\right)  ^{-}\right]  -\phi_{i}\left[  t\right]
\right)  \nonumber\\
&  =A\left[  t\right]  -A\left[  t-1\right]  \label{eq demand dolG}%
\end{align}
Similarly, for this system, the payoff to strategies based on the notional
wealth increase is given by Equation \ref{eq agent wealth chng} as:
\begin{align}
g_{R}\left[  t+1\right]   &  \varpropto\chi\left[  \phi_{R}\left[  t\right]
\Delta p\left[  t+1,t\right]  \right]  \nonumber\\
&  =\chi\left[  a_{R}^{\mu\left[  t-1\right]  }\left(  A\left[  t\right]
-A\left[  t-1\right]  \right)  \right]  \label{eq strat payoff posNWealth}%
\end{align}
\textbf{We will call this incarnation of the market model \$G\ref{eq strat
payoff posNWealth}}. The payoff structure for this one-position-per-agent
system is essentially the same as that reported in Equation [3] of `The
\$-Game' \cite{5dollar}.

\section{Behavior of market models\label{sec MM behavior}}

In the previous section, we introduced different microscopic reward schemes
for use in binary multi-agent market models. The aim of these new strategy
payoff structures, was to get the agents to maximize their notional wealth as
given by Equation \ref{eq agent wealth}. This seems to be a more physically
realistic goal than to have the agents competing to always be in the minority
group, as in the MG and its variants. The question now must be: how do these
new payoff structures affect the resultant dynamics of the market model? To
investigate this, we identify the form of the excess demand in the model and
then employ a price formation structure - from Section \ref{sec price process}
for example. The table below illustrates three possible market models:

\bigskip

\begin{center}%
\begin{tabular}
[c]{|l|l|l|l|l|}\hline
Name & $\phi_{R}\left[  t\right]  $ & $D\left[  \left(  t+1\right)
^{-}\right]  $ & $g_{R}\left[  t+1\right]  =\operatorname*{sgn}\left[
\ldots\right]  $ & $\Delta P\left[  t+1,t\right]  $\\\hline
GCMG & $\phi_{R}\left[  t-2\right]  +a_{R}^{\mu\left[  t-1\right]  }$ &
$A\left[  t\right]  $ & $-a_{R}^{\mu\left[  t\right]  }D\left[  \left(
t+1\right)  ^{-}\right]  $ & $D\left[  \left(  t+1\right)  ^{-}\right]
/\lambda$\\\hline
\$G\ref{eq strat payoff NWealth T} & $\phi_{R}\left[  t-2\right]  +a_{R}%
^{\mu\left[  t-1\right]  }$ & $A\left[  t\right]  $ & $\sum_{i=o}^{t-1}%
a_{R}^{\mu\left[  i\right]  }D\left[  \left(  t+1\right)  ^{-}\right]  $ &
$D\left[  \left(  t+1\right)  ^{-}\right]  /\lambda$\\\hline
\$G\ref{eq strat payoff posNWealth} & $a_{R}^{\mu\left[  t-1\right]  }$ &
$A\left[  t\right]  -A\left[  t-1\right]  $ & $a_{R}^{\mu\left[  t-1\right]
}D\left[  \left(  t+1\right)  ^{-}\right]  $ & $D\left[  \left(  t+1\right)
^{-}\right]  /\lambda$\\\hline
\end{tabular}
\end{center}

\bigskip

\begin{figure}
[ptb]
\begin{center}
\includegraphics[
natheight=9.030400in,
natwidth=7.661400in,
height=6.3495in,
width=5.3904in
]%
{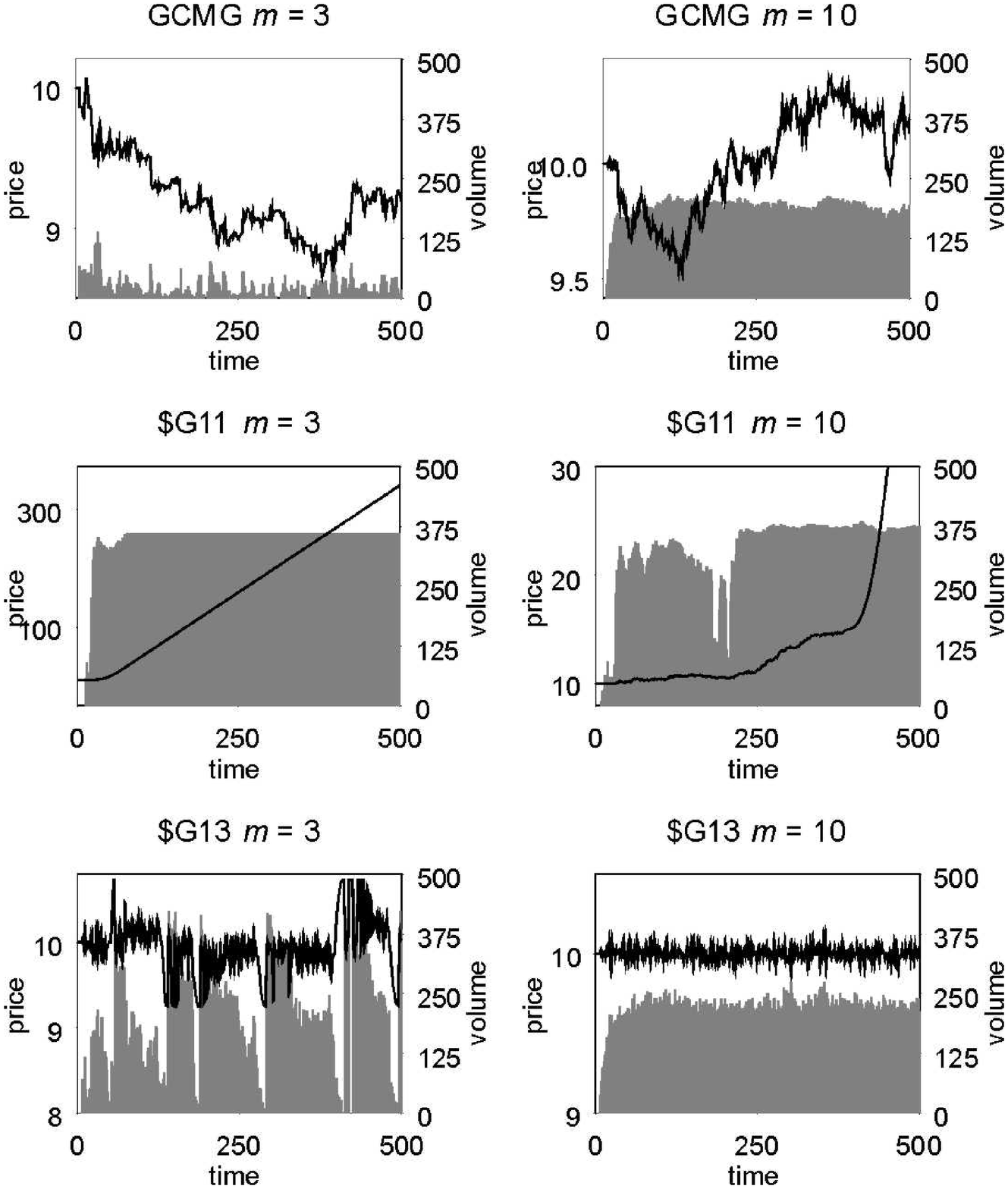}%
\caption{Behaviour of price and volume in market models corresponding to three
different strategypayoff functions $g_{R}\left[  t\right]  $. Parameters for
each model are $N=501$, $s=2$, $T=100$, $r=4$. The left hand column represents
the crowded market with high agent coordination ($m=3$) the right hand column
represents an uncrowded market with little agent coordination ($m=10$). The
top row shows the GCMG, the second row shows the \$G\ref{eq strat payoff
NWealth T} and the bottom row shows \$G\ref{eq strat payoff posNWealth}.}%
\label{fig MM price vol}%
\end{center}
\end{figure}

We have simulated the three market models in the above table with the same
parameters for each model: $N=501$ agents, $s=2$ strategies per agent, $T=100
$ timesteps for the strategy score time-horizon, and an $r=4$ payoff point
confidence-to-play threshold. Each simulation was run for agents with low
memory $m=3$ and for agents with high memory $m=10$. The resultant price
$p\left[  t\right]  $ and volume $v\left[  t\right]  =\sum_{R=1}^{2^{P}}n_{R}$
are displayed in Figure \ref{fig MM price vol}.%

The two different values of the agent's memory length (or rather the ratio
$P/N$) describe markets in two distinctly different `phases'. Arguably the
most important feature of these types of model is that there is a common
perception of a strategy's (virtual) success, i.e. $S_{R}\left[  t\right]  $
is a global variable. This implies that the number of agents adopting the same
action (i.e. the effect of crowding or herding) is dependent on the number
which hold each strategy. At low memory (low $P/N$) there are few strategies
available and many agents, consequently large numbers of agents hold the same
strategy. This implies that in this `crowded phase' there are large groups of
agents adopting the same action $a_{R}^{\mu\left[  t\right]  }$ (i.e. crowds)
and only small groups adopting the opposite action $-a_{R}^{\mu\left[
t\right]  }$ (i.e. anti-crowds) \cite{12crowd}. The result of this high agent
coordination is a volatile market with large asset price movements.
Conversely, for high $m$ (high $P/N$) hardly any agents hold the same
strategy. Hence the crowd size is roughly equal to the anti-crowd size,
resulting in a market with low agent coordination and consequently lower
volatility and fewer large changes. With this distinction between low and high
$m$ regimes in mind, let us briefly describe the dynamics in each market model.

\bigskip

\begin{itemize}
\item \textbf{GCMG}. First consider low $m$. The confidence-to-play threshold
is set high enough in order that occasional stochasticity is injected via
timesteps when $A\left[  t\right]  =0$ (N.B. the next state of $\mu\left[
t+1\right]  $ is then decided with a coin toss). The GCMG can be seen to
reproduce many of the statistical features observed in real markets. Figure
\ref{fig MM price vol} shows that the activity (volume) is generally low and
bursty. The asset price series is thus characterized by frequent large
movements (giving fat-tailed distributions of returns) and clustered
volatility. No long periods of correlation exist in the GCMG asset price
movement: as soon as lots of agents start taking the same action, then the
strategies which produce that action are penalized, as can be seen from
Equation \ref{eq strat payoff MG}. At high $m$, the absence of agent
co-ordination leads to an absence of clustering of activity. Hence the series
of asset prices appears more random at high $m$.

\item \textbf{\$G\ref{eq strat payoff NWealth T}}. At low $m$, the asset price
very soon assumes an unstoppable trend: agents with at least one
trend-following strategy ($\ldots A\left[  t-2\right]  >0,A\left[  t-1\right]
>0,A\left[  t\right]  >0\Rightarrow a_{R}=1$) join the trend and benefit
(notionally) from the consequent asset price movement. Because the strategies
(and hence agents) are allowed to accumulate limitless positions, the trend is
self-reinforcing since $\Delta W_{R}$\ just keeps getting bigger for the
trend-following strategies. At high $m$, the lack of agent coordination means
that it is harder for the model to find this attractor. However sooner or
later, a majority of trend-following strategies will have accumulated
sufficient score $S_{R}$ and position $\left|  \phi_{R}\right|  $ to be played
successfully. From then on, the pattern of success is self-reinforcing and
again the unstoppable trend is created. This result has to be seen as the
natural consequence of wanting the agents to maximize \emph{notional} wealth
at the same time as being allowed arbitrarily large positions.

\item \textbf{\$G\ref{eq strat payoff posNWealth}}. At low $m$, the model has
interesting and irregular dynamical properties. For example, the series of
asset price movements contains periods of very high correlation (and high
volume) and periods of very high anti-correlation (and lower volume). These
two behaviors arise from the payoff structure of Equation \ref{eq strat payoff
posNWealth}: just as in \$G\ref{eq strat payoff NWealth T}, strategies are
rewarded for having a positive (negative) position when the asset price is
trending up (down), however they now get penalized for \emph{joining} that
trend. This implies that persistent behavior will tend to follow persistent
behavior, while anti-persistent behavior will follow antipersistent behavior.
This finding is very much in line with the school of thought that says that
markets have distinct ranging and breakout phases. Unlike the \$G\ref{eq strat
payoff NWealth T} model, the trends are always capped by the limitation that
agents can only hold one position. Once as many agents as possible are long
(short) the asset price cannot rise (fall) any further: consequently the
resulting dynamics are strongly mean-reverting. At high $m$, due to the lack
of agent coordination, it is hard for large enough crowds to arise to form
trending periods. Thus anti-persistence dominates this regime.
\end{itemize}

\section{Instability in market models\label{sec Instabilities}}

In the MG, there is no unstable attractor which could give rise to a
continuously diverging price. The reason is that each agent is trying to do
the opposite of the others. This results in no net long-term market force. The
MG agents can see no advantage in collectively forcing a price up (down) to
increase the value of their long (short) position, since they are totally
unaware of their accumulated position. Section \ref{sec MM behavior} showed
however that, when we give the agents the goal of maximizing their notional
wealth (cash plus value of position), the model can be dominated by the
unstable dynamics of price trending. Although the structure of the \$G models
seems more realistic than the MG, the price dynamics do not reproduce many of
the stylized features of financial asset price series. It seems that an
improvement in the microscopic model structure `spoils' the macroscopic
results. To help explain this, we now look at some of the model deficiencies
which still remain.

\subsection{Market impact}

We hinted at this effect in Section \ref{sec MG to MM}. The wealth
$W_{i}\left[  t\right] $ that we have forced the agents of the \$G models to
maximize, is purely \emph{notional} - it is the wealth they \emph{would} have
if they could unwind their market positions at today's price $p\left[
t\right]  $. The agents are clearly not able to do this. There are two reasons
for this. Even if an agent put in an order to unwind his position at time $t$,
it would not be executed until time $t+1$ when the price is $p\left[
t+1\right]  $. The second reason is that in this simplistic structure, the
agents can only trade one quanta of asset at any time. Hence a position of
$\phi_{i}$ assets will take $\phi_{i}$ timesteps to fully unwind. All other
things being equal, an agent should expect on average to get less cash back
than $\phi_{i}\left[  t\right]  p\left[  t\right]  $ when unwinding his
position. This is `market impact'. We can see this effect directly with the
following argument:

\noindent The change in cash of agent $i$ from buying $\Delta\phi_{i}\left[
t,t-1\right]  $ assets at time $t$ is given by: $\Delta C_{i}\left[
t,t-1\right]  =-\Delta\phi_{i}\left[  t,t-1\right]  p\left[  t\right]  $. Let
us then use Equation \ref{eq price formation add} to give $p\left[  t\right]
=p\left[  t-1\right]  +D\left[  t^{-}\right]  /\lambda$. Then we break down
the demand as follows:
\[
D\left[  t^{-}\right]  =d_{i}+\sum_{j\neq i}^{N}d_{j}\left[  t^{-}\right]
\]
where $d_{j}$ is the demand of agent $j$. However we know that the demand of
agent $i$ is $\Delta\phi_{i}\left[  t,t-1\right]  $ assets, hence we get
\begin{align*}
\Delta C_{i}\left[  t,t-1\right]   &  =-\Delta\phi_{i}\left[  t,t-1\right]
p\left[  t\right] \\
&  =-\Delta\phi_{i}\left[  t,t-1\right]  p\left[  t-1\right]  -\frac
{1}{\lambda}\left(  \Delta\phi_{i}\left[  t,t-1\right]  \right)  ^{2}-\frac
{1}{\lambda}\Delta\phi_{i}\left[  t,t-1\right]  \sum_{j\neq i}^{N}d_{j}\left[
t^{-}\right]
\end{align*}
If we now assume that the remaining agents $j\neq i$ form an effective
`background' of random demands $d_{j}$ of zero mean, then averaging over this
`mean-field' of agents gives:
\begin{equation}
\left\langle \Delta C_{i}\left[  t,t-1\right]  \right\rangle _{j}=-\Delta
\phi_{i}\left[  t,t-1\right]  p\left[  t-1\right]  -\frac{1}{\lambda}\left(
\Delta\phi_{i}\left[  t,t-1\right]  \right)  ^{2}\label{eq mkt impact}%
\end{equation}
Equation \ref{eq mkt impact} tells us that an agent will on average receive an
amount of cash $\left\langle \Delta C_{i}\left[  t,t-1\right]  \right\rangle
$, which is equal to the amount he may have naively thought he would get (if
he could have traded at price $p\left[  t-1\right]  $) minus a positive amount
proportional to the square of his trade size and inversely proportional to the
market liquidity. Consequently, an agent who considered his market impact in
calculating his wealth would on average expect the cash-equivalent value of
his wealth to be given by $W_{i}\left[  t\right]  -\phi_{i}\left[  t\right]
^{2}/\lambda$, i.e. his notional wealth minus the average impact from
unwinding his position at time $t+1$. This is the wealth that the agents
\emph{should} be attempting to maximize. We could therefore propose, in the
spirit of Ref. \cite{13marsili}, a modification to the strategy reward
function $g_{R}\left[  t\right]  $ which takes account of this market impact:
\begin{equation}
g_{R}\left[  t+1\right]  \varpropto\chi\left[  \phi_{R}\left[  t\right]
\Delta p\left[  t+1,t\right]  -\phi_{R}\left[  t\right]  ^{2}/\lambda\right]
\label{eq strat payoff Impact}%
\end{equation}

Let us now re-visit the issue of trend following, this time with reference to
a model whose strategies are updated using Equation \ref{eq strat payoff
Impact}. \textbf{We will refer to this model as \$G\ref{eq strat payoff
Impact}}. When a trend forms such that $\phi_{R}\left[  t\right]  $ and
$\Delta p\left[  t+1,t\right]  $ have the same sign, strategy $R$ will be
rewarded: this continues until the moment when the extra profit gained from
increasing the position to benefit from the favorable price movement, is
smaller than the added market impact which would be faced from unwinding the
larger position. At this point the strategy will start to be penalized instead
of rewarded for supporting the trend. This is a mechanism which could halt the
price-divergent behavior of \$G\ref{eq strat payoff NWealth T} type models.
However Equation \ref{eq strat payoff Impact} gives the turnover point in
strategy reward due to market impact, to be given by $\phi_{R}\left[
t\right]  =\lambda\Delta P\left[  t+1,t\right]  $. If we then consider that
during an `unstable' market trend the average price change is of order
$N/\lambda$ (see, for example, the gradient of the \$G\ref{eq strat payoff
NWealth T} price in Figure \ref{fig MM price vol}) - and that $\phi_{R}$
increments by only $\pm1$ every timestep - then trends which are formed before
the market impact mechanism starts having a strong effect, will be of order
$t_{\text{trend}}\backsim N$ timesteps in length. The market impact mechanism
on its own, is thus clearly too weak to stabilize the \$G models in order to
form a realistic price series.

\subsection{Agent wealth}

Financial agents participating in a real market have finite resources and so
cannot keep buying and/or selling assets indefinitely. This hard cut-off of
agents' resources in turn imposes a hard limit on the magnitude of price
trends. This effect was seen clearly in the \$G\ref{eq strat payoff
posNWealth} model where the agents were only permitted (or, equivalently
because of finite resources, were only able to) hold one position at any one
time. The \$G\ref{eq strat payoff posNWealth} and \$G\ref{eq strat payoff
NWealth T} models can thus be thought of as opposite extremes in this respect,
i.e. the first mimics extremely tightly-limited resources while the second
mimics infinite resources. It seems natural to expect that a real financial
market lies somewhere between these two extremes; resources are large for the
market-moving agents but still finite. We can include the effect of limited
agent resources in our market model by allocating agent $i$ an initial capital
$C_{i}\left[  0\right]  $ (and position $\phi_{i}\left[  0\right]  $) and then
updating this capital using $C_{i}\left[  t\right]  =C_{i}\left[  t-1\right]
-\Delta\phi_{i}\left[  t,t-1\right]  p\left[  t\right]  $ \cite{4liege}. The
agents cannot then trade at time $t$ if the following conditions are met:
\begin{align*}
\Delta\phi_{i}\left[  t,t-1\right]   &  >0\cap C_{i}\left[  t-1\right]
<\Delta\phi_{i}\left[  t,t-1\right]  p\left[  t-1\right] \\
\Delta\phi_{i}\left[  t,t-1\right]   &  <0\cap\phi_{i}\left[  t-1\right]  <0
\end{align*}
In other words, an agent will not submit a buy order unless he at least has
the capital to buy the asset at the quoted price $p\left[  t-1\right] $, and
also will not submit an order to short sell if he already holds a short
position. If we imposed no limit on short selling, an unstable state of the
system would exist wherein all agents short-sell indefinitely.

\begin{figure}
[ptb]
\begin{center}
\includegraphics[
natheight=5.955100in,
natwidth=7.726200in,
height=3.5725in,
width=4.6363in
]%
{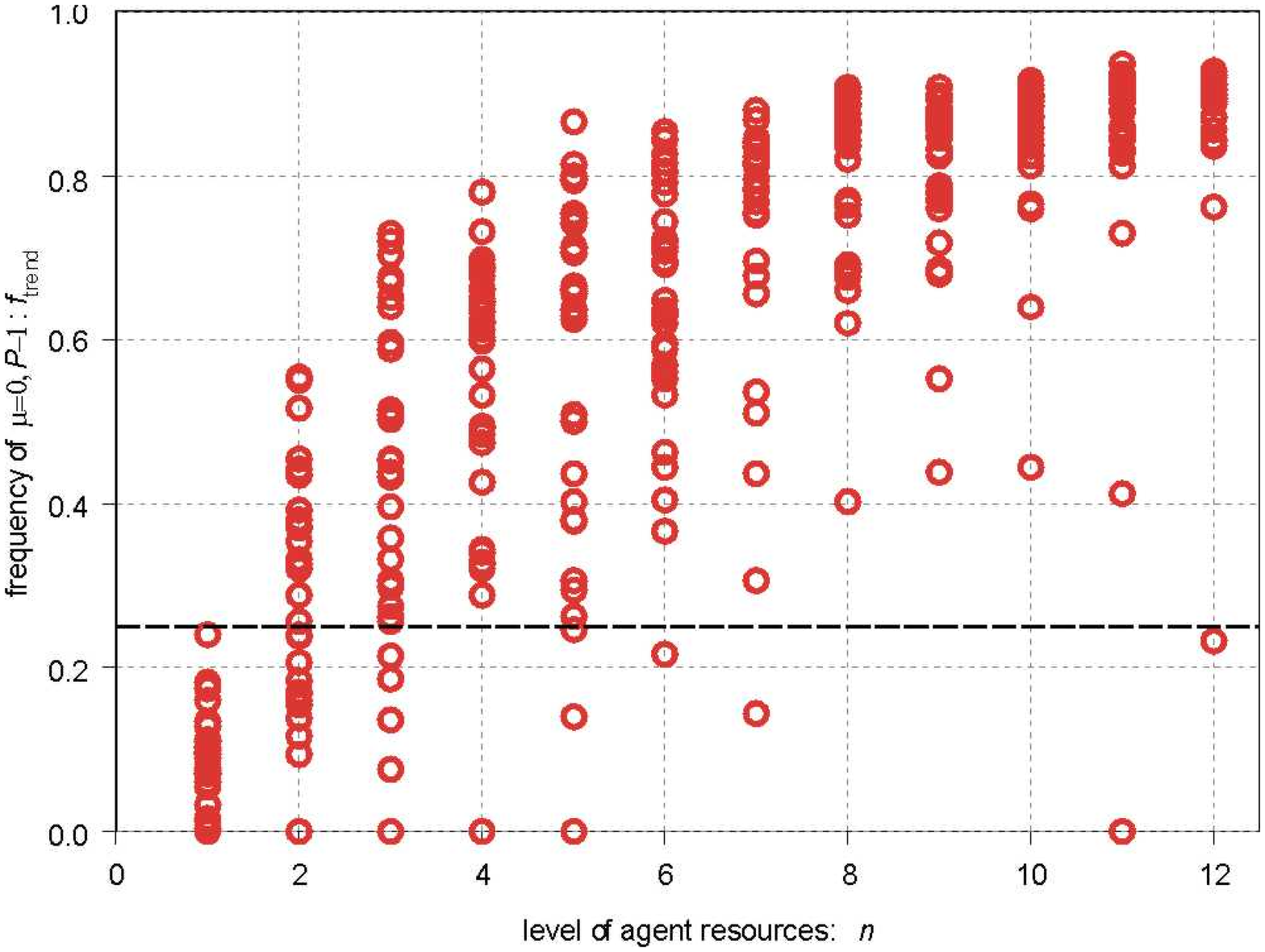}%
\caption{The frequency of occurence of the global information states
$\mu=0,P-1$, $f_{\text{trend}}$, as a function of the agents' capital resource
level $n$. The results were taken from the last 500 timesteps of a 1500
timestep simulation. The market model used was \$G\ref{eq strat payoff NWealth
T}W, with evolving agent wealths. The model parameters were $N=501$, $m=3$,
$s=2$, $T=100$, and $r=4$, with binary ($\chi=\operatorname*{sgn}$) strategy
rewards. The solid line represents equally visited $\mu$-states.}%
\label{fig Ftrend}%
\end{center}
\end{figure}

\textbf{We will refer to this model as \$G\ref{eq strat payoff NWealth T}W}
where W represents `wealth'. Let us initialize this market model with agent
wealths such that the agents' initial buying power is equal to their initial
selling power, i.e. $\left\{  C_{i}\left[  0\right]  ,\phi_{i}\left[
0\right]  \right\}  =\left\{  n\times p\left[  0\right]  ,n-1\right\}  .$
\ Hence initially, the agents have the power to buy or sell $n$ assets (N.B.
recall that they are allowed up to one short sell). If we then raise $n$ from
$n=1$ upwards, we can see a qualitative change in the dynamical behavior of
the model between the two extremes of \$G\ref{eq strat payoff posNWealth} and
\$G\ref{eq strat payoff NWealth T}, with the periods of trending growing
longer as $n$ increases. To investigate this behavior numerically, we fix $n$
and run the model for 1500 timesteps, recording the value of the global
information $\mu\left[  t\right]  $ in the last 500 timesteps $t$. We then
count the number of times within this last 500 timesteps that either $\mu=0$
(implying negative price movements over the last $m$ timesteps) or $\mu=P-1$
(positive price movements) occur. We denote the frequency with which these
states occur as $f_{\text{trend}}$. If the model visited all states of the
global information $\mu\left[  t\right]  $ equally, we would expect
$f_{\text{trend}}=1/P$. Figure \ref{fig Ftrend}, showing the variation of
$f_{\text{trend}}$ with $n$, has several interesting features. First it can be
seen that, as $n$ is increased and the consequent wealth available to the
agents grows, the tendency of the model to be dominated by price trending
increases dramatically. Also we see that at low $n$, $f_{\text{trend}}$ is
below that of the (random) equally visited $\mu$-states case. This is due to
the high degree of anti-persistence in the system as was seen in the
\$G\ref{eq strat payoff NWealth T} model. The large spread in the results
arises from the tendency of the model to exhibit clustering of activity
states: persistence follows persistence and anti-persistence follows
antipersistence as described in Section \ref{sec MM behavior}.%

We note that in the work of Giardina and Bouchaud (see Refs.
\cite{6bouchaudlong,9bouchaudprague}) the strategy payoff function has the
same basic form as Eq. \ref{eq strat payoff posNWealth}, i.e. it is a
\$G\ref{eq strat payoff posNWealth}W-style model. These authors consider
proportional scoring (i.e. $\chi=1$) and include an interest rate which
resembles a resource level $L$; these features will not change the qualitative
results presented here. However, in the models of Refs.
\cite{6bouchaudlong,9bouchaudprague}, it seems that only active strategies are
rewarded thereby breaking the common perception of a given strategy's success
among agents which is arguably a central feature of the binary model
structure. We expect this to dramatically affect the dynamical properties of
the system as the mechanism for agent coordination and thus crowd formation
has been altered.

\subsection{Diversity and timescales}

The interesting dynamical behavior of market models based on the binary
strategy approach of the MG, arises largely as a consequence of the
heterogeneity in the strategies held by the agents and the common perception
of a strategy's success. The extent to which agents will agree on the best
course of action, and the consequent crowding, is solely a function of how the
$2^{P}$ available strategies are distributed amongst the population of agents.
If the way in which the strategies are initially allocated is biased, then the
resulting dynamics will reflect this \cite{6bouchaudlong}.

\begin{figure}
[ptb]
\begin{center}
\includegraphics[
natheight=5.955100in,
natwidth=7.726200in,
height=4.4936in,
width=5.8228in
]%
{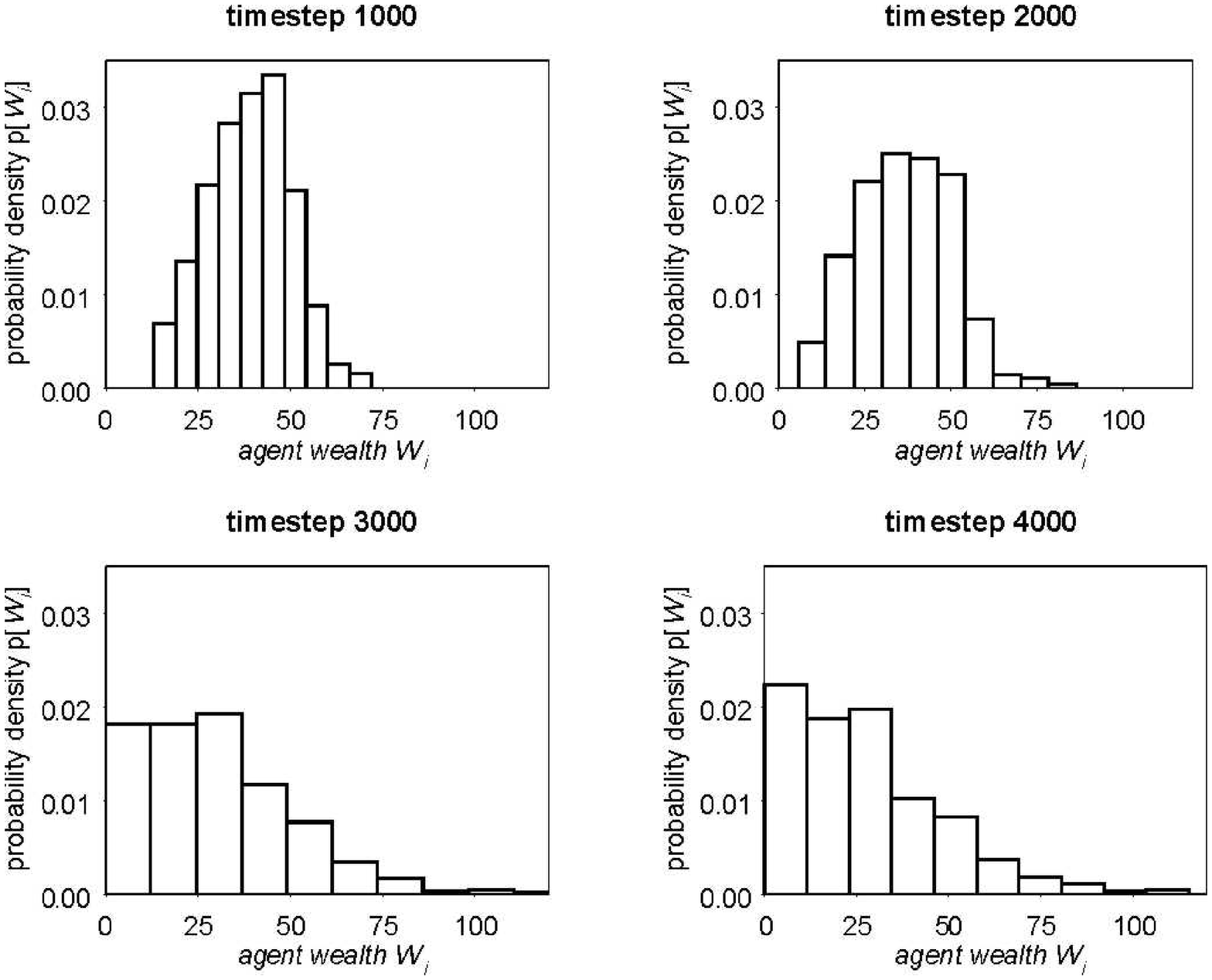}%
\caption{Distributions of the agents' wealth $W_{i}$ at four different times
$t$ during evolution of a \$G\ref{eq strat payoff NWealth T}W model. Initially
all agents were allocated the same resources $\left\{  C_{i}\left[  0\right]
,\phi_{i}\left[  0\right]  \right\}  =\left\{  3p\left[  0\right]  ,2\right\}
$ with the initial price $p\left[  0\right]  =10$. Parameters for the
simulation were $N=1001$, $m=3$, $s=2$, $T=100$, and $r=4$, with binary
($\chi=\operatorname*{sgn}$) strategy rewards.}%
\label{fig wealth PDFs}%
\end{center}
\end{figure}

The introduction of an agent wealth, as discussed in the previous subsection,
also brings about diversity in the market model. Even if we initiate the model
with all agents having an equal allocation of wealth in the form of cash plus
assets, the wealths of the agents $\{W_{i}\}$ will soon become heterogenous as
a direct result of their heterogenous strategies. Figure \ref{fig wealth PDFs}
shows the heterogeneity of agents' wealth growing with time during a
\$G\ref{eq strat payoff NWealth T}W simulation. After many timesteps have
elapsed, the distribution of agents' wealth seems to reach a steady-state:
many agents have lost the majority of their wealth to a minority of agents who
themselves have accumulated much more. In everyday terminology, `the rich get
richer while the poor get poorer'.%

The heterogeneity of agents' wealth in the \$G\ref{eq strat payoff NWealth T}W
model is fed back into the system through the buying power of the agents
alone. However, although wealthier agents have the potential to buy and sell
more assets, they still only trade in single quanta of the asset at any given
timestep, the same as poorer agents. It may be more reasonable to propose that
the agents trade in sizes proportional to their resources
\cite{4liege,6bouchaudlong}, i.e.
\begin{align*}
\Delta\phi_{i}\left[  t,t-1\right]   &  =\gamma\frac{C_{i}\left[  t-1\right]
}{p\left[  t-1\right]  }\text{ \ \ for \ \ }a_{R_{i}^{\ast}}^{\mu\left[
t-1\right]  }=1\\
\Delta\phi_{i}\left[  t,t-1\right]   &  =\gamma\phi_{i}\left[  t-1\right]
\text{ \ \ for \ \ }a_{R_{i}^{\ast}}^{\mu\left[  t-1\right]  }=-1
\end{align*}
where $R_{i}^{\ast}$ is the highest scoring strategy $R$ of agent $i$. The
factor $\gamma$ then enumerates what fraction of the agents' resources (cash
for buying, assets for selling) they will transact at any given time. In this
system, assets will generally need to be divisible, i.e. the sense of agents
buying and selling a quanta of the asset as in \$G\ref{eq strat payoff NWealth
T}W is lost. This in turn means that instead of the degree of trending being
controlled by the level of initial resource allocation $n$, it is instead
determined by $n/\gamma$, since this effectively determines the number of
trades agents can make in any trending period before hitting the boundary of
their capital resources. Also with this system of trading in proportion to
wealth, trends will start steep and end shallow as agents run out of resources
and thus make smaller and smaller trades. Apart from these stylized
differences, the system has a very similar dynamical behavior to the more
straight-forward \$G\ref{eq strat payoff NWealth T}W model.

\begin{figure}
[ptb]
\begin{center}
\includegraphics[
natheight=3.491200in,
natwidth=7.726200in,
height=2.2978in,
width=5.0505in
]%
{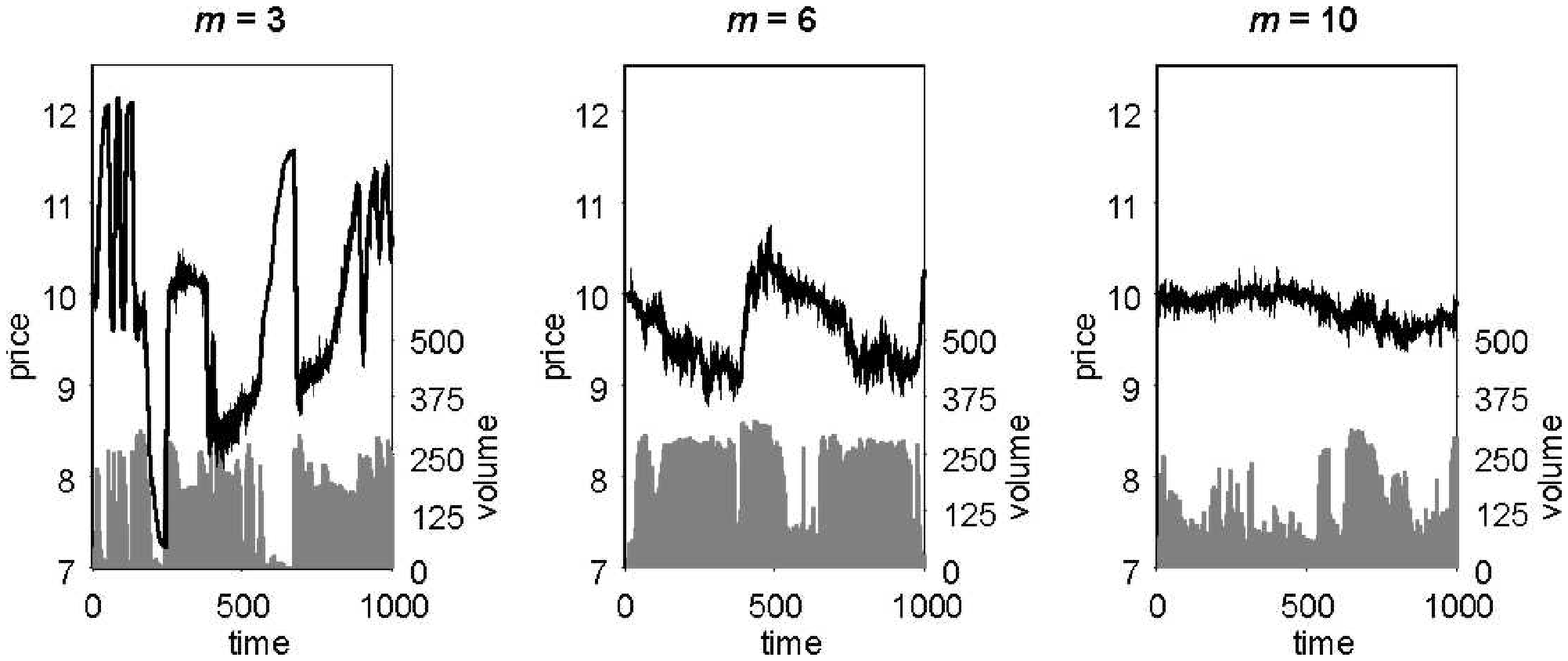}%
\caption{Examples of the dynamical behavior of the price $p\left[  t\right]  $
and volume $\sum_{i=1}^{N}\left|  d_{i}\right|  $ of the \$G\ref{eq strat
payoff NWealth T}W model, for three different degrees of crowding given by the
memory length $m$. The parameters of the simulation are $N=501$, $s=2$,
$T=100$, $r=4$, and $n=3$, with binary ($\chi=\operatorname*{sgn}$) strategy
rewards.}%
\label{fig dolGW price vol}%
\end{center}
\end{figure}

Diversity in strategies and wealth are the two big sources of agent
heterogeneity that we have covered so far. This agent heterogeneity has led to
a market model with dynamical behavior which is interesting and diverse over a
large parameter range. However, the typical price/volume output only starts to
become representative of a real financial market at higher $m$, as can be seen
in Figure \ref{fig dolGW price vol}. As discussed in Section \ref{sec MM
behavior}, the high $m$ regime represents a `dilute' market where very few
traders act in a coordinated fashion. However one would expect that a real
financial market is \emph{not} in a dilute phase at all, since it \emph{does}
have large groups of agents forming crowds which rush to the market together,
thereby creating the bursty activity pattern observed empirically. Why then
does this model, when pushed into the low $m$ regime, produce endless bubbles
of positive and negative speculation as shown in the left-hand panel of Fig.
\ref{fig dolGW price vol}. Although a real financial market does show
\emph{some} suggestion of oscillatory bubble formation, it is nowhere near so
pronounced and is certainly not on such a short timescale\footnote{We assume
implicitly here that a `timestep' in our model corresponds to a fairly short
interval in real time. The reason for this is that a timestep is the amount of
time it takes for an agent to re-assess his strategies. For large, market
moving, agents this amount of time is likely to be less than one day.}.%

To answer this question, we push further the subject of agent diversity.
Although our agents may have differing sets of strategies and consequently
different wealths, they all act on the same timescale. When we look at charts
such as Figure \ref{fig dolGW price vol}, we see patterns not only on a small
tick-by-tick scale but also on a much larger scale. In fact we can identify
patterns all the way up to the `macro' scale of the boom-bust speculatory
bubbles. From a knowledge of these patterns, we would form opinions about what
will happen next and would trade accordingly to maximize our wealth. The
agents we have modelled, however, cannot view the past price-series in this
way; instead they are forced to only consider patterns of length $m$
timesteps. Patterns of any length greater than this, go un-noticed by the
agents and hence are not traded upon. This explains why these patterns can
arise and survive. By contrast, when a pattern is traded upon (even in the
non-MG framework) it is slowly removed from the market. Consider the following pattern:

\begin{center}%
\begin{tabular}
[c]{lllllllllllllll}%
time $t$ & 1 & 2 & 3 & 4 & 5 & 6 & 7 & 8 & 9 & 10 & 11 & 12 & 13 & 14\\
price $p\left[  t\right]  $ & 14 & 10 & 11 & 12 & 13 & 15 & 14 & 10 & 11 &
12 & 13 & 15 & 14 & 10
\end{tabular}
\end{center}

If we were able to identify this pattern repeating, our best course of action
would be to submit a buy order between times $t=1$ and $t=2$, i.e. $a\left[
1\right]  =1$. We could then buy the asset at $p\left[  2\right]  =10$. Then
between times $t=5$ and $6$, we would submit an order to sell: $a\left[
5\right]  =-1$ and hence would have sold the asset at $p\left[  6\right]
=15$. We then continue: $a\left[  7\right]  =1$, $a\left[  11\right]  =-1$...
This ensures we always buy at the bottom price and sell at the top. However,
trading in this way is \emph{against} the trend as $a\left[  t\right]
A\left[  t\right]  <0$. It is in effect minority trading. Hence trading to
maximize our wealth with respect to this pattern, leads to the weakening of
the pattern itself just as in the MG. We conclude therefore that the presence
of strong patterns in the \$G\ref{eq strat payoff NWealth T}W and other
similar market models at low $m$, is simply due to the absence of agents
within the model who can identify these patterns and arbitrage them out.

From the proceeding discussion, it is clear that a realistic market model
should include agents who can analyze the past series of asset price movements
over different time-scales. A first thought on how to do this, is to include a
heterogeneity in the memory length $m$ of the agents. In this framework,
agents look at patterns not only of differing length but also of differing
complexity. It is therefore appealing to propose a generalization to the way
in which the agents interpret the global information of past price movements,
in such a way as to allow observation of patterns over different times-scales
but having the \emph{same} complexity ($m$). This can be simply achieved by
allowing the agents to have a natural information bit-length $\tau$ such that
the global information available to them $\mu_{\tau}\left[  t\right]  $ is
updated according to the following generalization of Equation \ref{eq history
update Gen}:
\[
\mu_{\tau}\left[  t+1\right]  =2\mu_{\tau}\left[  t\right]  -P\operatorname{H}%
\left[  \mu_{\tau}\left[  t\right]  -\frac{P}{2}\right]  +\operatorname{H}%
\left[  \Delta P\left[  t+1,t+1-\tau\right]  \right]
\]
The table below then shows how the previous pattern would be encoded by agents
having $\tau=1,2,3,4$:

\begin{center}%
\begin{tabular}
[c]{|l|l|l|l|l|l|l|l|l|l|l|l|l|l|l|}\hline
time $t$ & $1$ & $2$ & $3$ & $4$ & $5$ & $6$ & $7$ & $8$ & $9$ & $10$ & $11$ &
$12$ & $13$ & $14$\\\hline
price $p\left[  t\right]  $ & $14$ & $10$ & $11$ & $12$ & $13$ & $15$ & $14$ &
$10$ & $11$ & $12$ & $13$ & $15$ & $14$ & $10$\\\hline
`best' action $a^{\ast}\left[  t\right]  $ & $1$ & $1$ & $1$ & $1$ & $-1$ &
$-1 $ & $1$ & $1$ & $1$ & $1$ & $-1$ & $-1$ & $1$ & $1$\\\hline
$\operatorname*{sgn}\left[  \Delta P\left[  t,t-1\right]  \right]  $ &  & $-$%
& $+$ & $+$ & $+$ & $+$ & $-$ & $-$ & $+$ & $+$ & $+$ & $+$ & $-$ &
$-$\\\hline
$\operatorname*{sgn}\left[  \Delta P\left[  t,t-2\right]  \right]  $ &  &  &
$-$ & $+$ & $+$ & $+$ & $+$ & $-$ & $-$ & $+$ & $+$ & $+$ & $+$ & $-$\\\hline
$\operatorname*{sgn}\left[  \Delta P\left[  t,t-3\right]  \right]  $ &  &  &
& $-$ & $+$ & $+$ & $+$ & $-$ & $-$ & $-$ & $+$ & $+$ & $+$ & $-$\\\hline
$\operatorname*{sgn}\left[  \Delta P\left[  t,t-4\right]  \right]  $ &  &  &
&  & $-$ & $+$ & $+$ & $-$ & $-$ & $-$ & $-$ & $+$ & $+$ & $-$\\\hline
\end{tabular}
\end{center}

If each agent only considered the past two bits of information, i.e. $m=2$,
then the `best' strategies for different values of the information bit-length
$\tau$ would be as shown in the table below. These `best' strategies have been
obtained from inspection of the table above; in particular, by looking at when
the different $m=2$ bitstrings $\left\{  --,-+,+-,++\right\}  $ occur and
seeing the respective `best' action given this bit string $a^{\ast}\left[
t\right]  $.

\begin{center}%
\begin{tabular}
[c]{cc}%
& $\tau$\\
$\mu\left[  t\right]  $ &
\begin{tabular}
[c]{|l|l|l|l|l|}\hline
& $1$ & $2$ & $3$ & $4$\\\hline
$--$ & $1$ & $1$ & $1$ & $?$\\\hline
$-+$ & $1$ & $1$ & $-1$ & $-1$\\\hline
$+-$ & $1$ & $1$ & $1$ & $1$\\\hline
$++$ & $?$ & $?$ & $?$ & $1$\\\hline
\end{tabular}
\end{tabular}
\end{center}

The `best' strategies ($\underline{a}_{R^{\ast}}$) in the above table, show a
question mark (?) next to a particular value of the global information
$\mu\left[  t\right] $. This denotes that for this state, the best action is
sometimes $a_{R^{\ast}}^{\mu\left[  t\right]  }=1$ and sometimes $a_{R^{\ast}%
}^{\mu\left[  t\right]  }=-1$. It can thus be seen that only when we include
longer time-scale patterns $\tau>2$, do we get a clear signal of when is the
optimal time to sell ($a_{R^{\ast}}^{\mu\left[  t\right]  }=-1$). Shorter
timeframe patterns provide no clear indication when to do so\footnote{In this
case a single $m=3$ strategy could encode enough information to optimally
arbitrage the pattern. However in general, for longer patterns, this is less
likely to be true.}. This then demonstrates that an agent holding strategies
of different bit-length $\tau$ could identify optimal times to buy and sell,
and hence arbitrage patterns of length very much greater than the memory
length $m$.

\subsection{Exogenous information}

So far this section has discussed how instabilities and inefficiencies in the
model market (as indicated by repeating, un-arbitraged patterns) could be
remedied by inclusion of more realistic, sophisticated and diverse strategic
agents. However the agents we have considered have all acted in the same
fashion, i.e. with the goal of maximizing their wealth by employing a
strategic methodology based on the observable endogenous market indicators. In
terms of a real financial market, we have considered modelling the subset of
agents who consider the past history of asset price movements as the only
relevant information. Although in reality we expect this subset to be large,
we certainly acknowledge the presence of agents who strategically use other
endogenous market indicators such as volume as a source of global information.
The strategic use of other endogenous market indicators simply gives a more
sophisticated model within the same basic framework.

What we have so far neglected to include is the presence of agents using
exogenous information to influence their trading decisions. Possible sources
of exogenous information are as vast as human imagination itself: they could
for example range from economic indicators through to company reports and
rumor, or even `gut-feeling' and astrology! Within the context of our model
framework, strategies employed based on these external sources of information
are un-strategic with respect to the endogenous information, and as such
represent a stochastic influence.

Such stochasticity, representing the response of agents to exogenous
information, can be incorporated into the models in a variety of ways. For
example a background level of stochastic action could be assumed at each
timestep, or be made to appear as a Poisson process with a given mean
frequency. However, it seems most physically satisfying to incorporate
stochastic action in the form of an extra pair of strategies for each (or at
least some) agent: one buy strategy, $\underline{a}_{R}=\underline{1}$, and
one sell strategy, $\underline{a}_{R}=-\underline{1}$. Agents could then be
given a probability of using each of these strategies in place of their
`usual' chartist strategies. Furthermore, the probabilities of using these
strategies could be allowed to evolve. This system of modelling response to an
exogenous signal is essentially the `Genetic Game' of Ref. \cite{14genetic}.
Alternatively, the probability of using one of these strategies representing
the response to an exogenous signal, could be based on an endogenous market
indicator. An example of this is the mechanism of Ref. \cite{6bouchaudlong} of
switching to `fundamentalist behavior'. In the model of Ref.
\cite{6bouchaudlong}, agents use strategy $\underline{a}_{R}=\underline
{\operatorname*{sgn}\left[  p_{0}-p\left[  t\right]  \right]  }$ with a
probability $f$ which increases with $\left|  p\left[  t\right]  -p_{0}\right|
$, where $p_{0}$ represents a perceived `fundamental' value for the asset.
This mechanism, which can be thought of as encoding the behavior of irrational
fear, clearly acts to break the formation of long trends since the probability
of trading against the trend increases with the duration of the trend itself.
More generally, we suspect that among the agent-based market models in the
literature which manage to reproduce stylized facts, the inclusion of some
level of stochasticity within the model is crucial for helping to break up
unphysical dynamics and hence achieve market-like behavior.

\section{Strategy structure\label{sec strat struc}}

In the framework of the MG, each strategy $\underline{a}_{R}$ comprises the
elements $a_{R}^{\mu}$ which represent a response $\left\{  -1,1\right\}  $ to
each of the possible $P$ values of the global information $\mu$ (see Section
\ref{sec MG basics}). In Section \ref{sec MG to MM}, we identified the
different states of the global information $\mu$ as corresponding to different
patterns in the past history of asset price movements (binary: up/down). We
can therefore think of a strategy as a book of chartist principles
recommending an action for each and every possible pattern of length $m$. Each
of these strategy `books' thus has $2^{m}$ pages, one for each pattern. There
are $2^{2^{m}}$ possible books an agent can buy and use as guidelines on how
to trade. An agent in possession of one of these strategy `books', whichever
he chooses to buy, can find a page giving guidelines on how to trade in
\emph{every} possible market state. The books are thus `complete'.

It seems unlikely however, that in reality such a `complete' book would exist
for any arbitrary value of $m$. Indeed, even if such a book did exist, it is
unlikely that a given market participant would consider all the pages as being
true. For example, page one of a strategy book may say that if the asset price
has fallen three days running, one should sell assets on the fourth day. Page
three may say that if the asset price has fallen, then recovered and then
fallen again, it is advisable to buy. The agent holding this book may well
believe in page one but think that the guidelines of page three are rubbish:
in particular, he considers the pattern down-up-down to correspond to no
trading signal at all. This agent would therefore continue to hold any
previous position he had if he saw the pattern down-up-down. We therefore
propose a generalization to the strategy structure of the MG to account for
the fact that strategy `books' may be incomplete, or equivalently that agents
do not trust some of the trading signals.

Before trading commences, each agent goes through each of her (complete)
strategy `books' page by page, and decides whether she believes in the trading
guidelines given for each of the possible $P=2^{m}$ market `signals'. The
agents choose to accept each page of each of their books with a probability
$p_{a}$. If the agent rejects a page of the strategy `book', they replace the
suggested action for time $t$, i.e. $a_{R}^{\mu\left[  t\right]  }\in\left[
-1,1\right]  $, by a null action $a_{R}^{\mu\left[  t\right]  }=\ast$. When
trading commences, the agents are then faced with deciding what course of
action to take when their strategy `books' register the null action at time
$t$. One type of possible behavior would be to maintain their prior course of
action, i.e. $a_{R}^{\mu\left[  t\right]  }=a_{R}^{\mu\left[  t-1\right]  }$.
Another alternative behavior would be to maintain their prior position, i.e.
$\phi_{i}\left[  \left(  t+1\right)  ^{-}\right]  =\phi_{i}\left[  t\right] $.
(This results in $a_{R}^{\mu\left[  t\right]  }=0$ in \$G\ref{eq strat payoff
NWealth T} type models and $a_{R}^{\mu\left[  t\right]  }=a_{R}^{\mu\left[
t-1\right]  }$ in \$G\ref{eq strat payoff posNWealth} models).

\begin{figure}
[ptb]
\begin{center}
\includegraphics[
natheight=6.664200in,
natwidth=7.338800in,
height=3.96in,
width=4.3578in
]%
{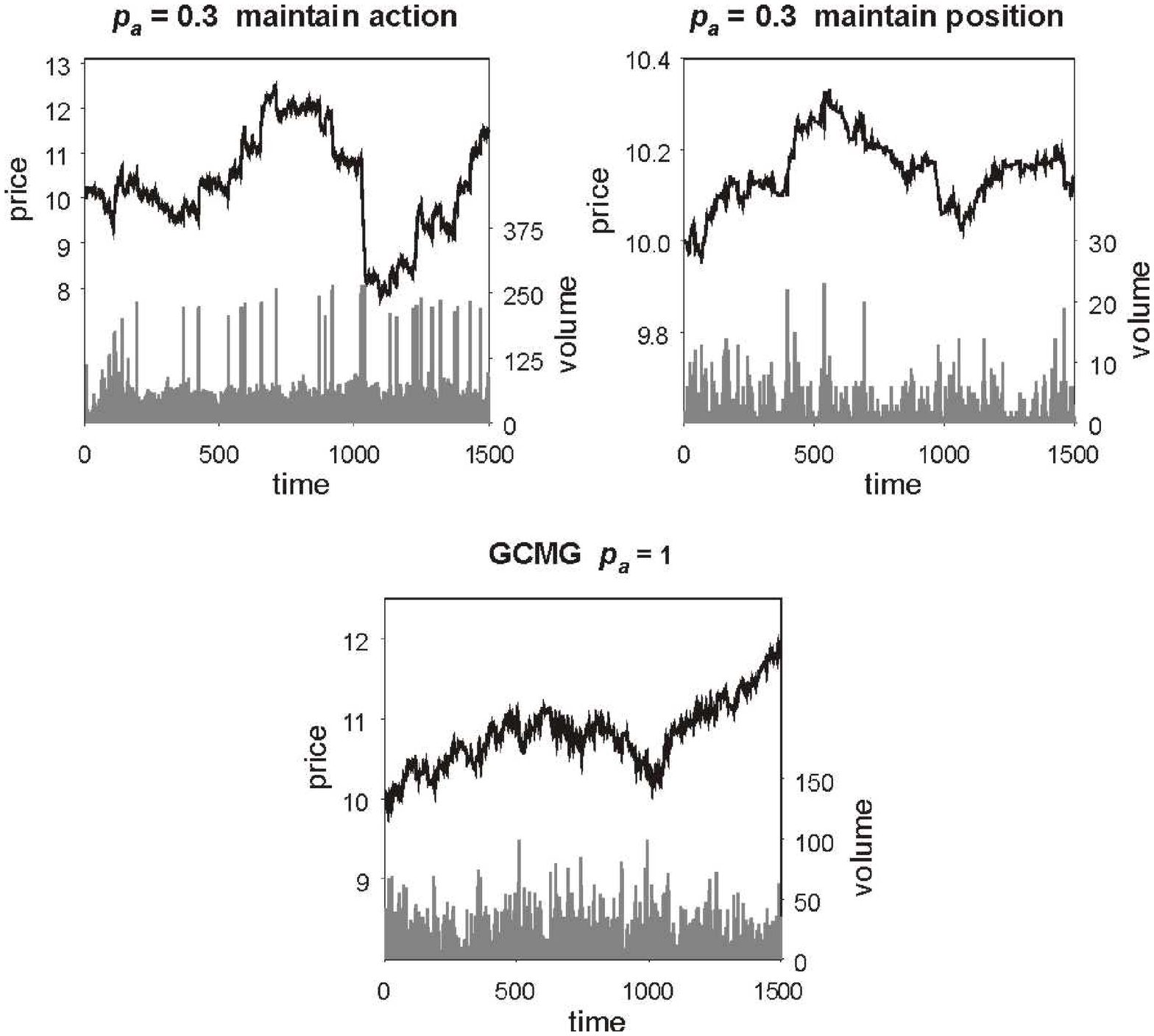}%
\caption{The contrast between the behaviour of models with incomplete strategy
`books' to those with complete ones. The top two figures show price and volume
for a GCMG type model with incomplete strategies, $p_{a}=0.3$. The top left
chart represents a market of agents who maintain their previous action when
they encounter a null strategy response $a_{R}^{\mu\left[  t\right]  }=\ast$.
The top right chart represents a market of agents who maintain their position.
The bottom chart shows the behavior of the GCMG for comparison. The parameters
in all the models are $N=501$, $s=2$, $T=100$, and $r=4$, with binary
($\chi=\operatorname*{sgn}$) strategy rewards.}%
\label{fig ternGW price vol}%
\end{center}
\end{figure}

We now examine the qualitative differences between a model with this
incomplete strategy structure, and a model with complete strategies. For this
purpose we will assume a GCMG type model, i.e. a model with strategy payoff
given by Equation \ref{eq strat payoff MG}\footnote{We use the GCMG to
contrast the effects of incomplete strategies, since its dynamics are familiar
and have been well-studied. However our previous comments regarding the
shortcomings of the MG strategy payoff structure, should be kept in mind since
they still hold.}. Figure \ref{fig ternGW price vol} shows the resulting price
and volume from two market simulations with incomplete strategies, the
probability of non-null action being $p_{a}=0.3$. These are compared with the
GCMG which has complete strategies.%

The top-left chart of Figure \ref{fig ternGW price vol} shows price and volume
for a market where agents maintain their previous course of action when they
encounter a null strategy action. This market shows high volatility in price,
a high number of very large movements, and a background of high volume trading
with occasional sharp spikes. We suggest that this behavior arises from the
tendency of agents to keep repeating their previous action in the absence of a
new market pattern. Clearly then the `large-movement' states of up-up-up or
down-down-down will be occupied for longer than the `ranging' states
(up-down-up etc.) because these are the only states of $\mu$ which can map
onto themselves ($\mu\left[  t\right]  =\mu\left[  t+1\right]  =\mu\left[
t+2\right]  \ldots$) thus providing no new market pattern.

The top right chart of Figure \ref{fig ternGW price vol} shows price and
volume for a market where agents maintain their previous position (i.e.
$a_{R}^{\mu\left[  t\right]  }=0$ in this model) when they encounter a null
strategy action. The volume in this market simulation is very low, giving a
consequent low volatility. This is due to the fact that the agents will not
trade unless they receive a market signal for which they have a non-null
strategy action. This model still shows a high probability for price movements
which are large compared to the volatility. These large movements occur in
moments of high agent coordination, i.e. when many agents find they have a
strategy with the same non-null action. However in contrast to the previous
model, these large movements don't seem to persist over many timesteps.

\begin{figure}
[ptb]
\begin{center}
\includegraphics[
natheight=4.200400in,
natwidth=7.726200in,
height=2.7579in,
width=5.0505in
]%
{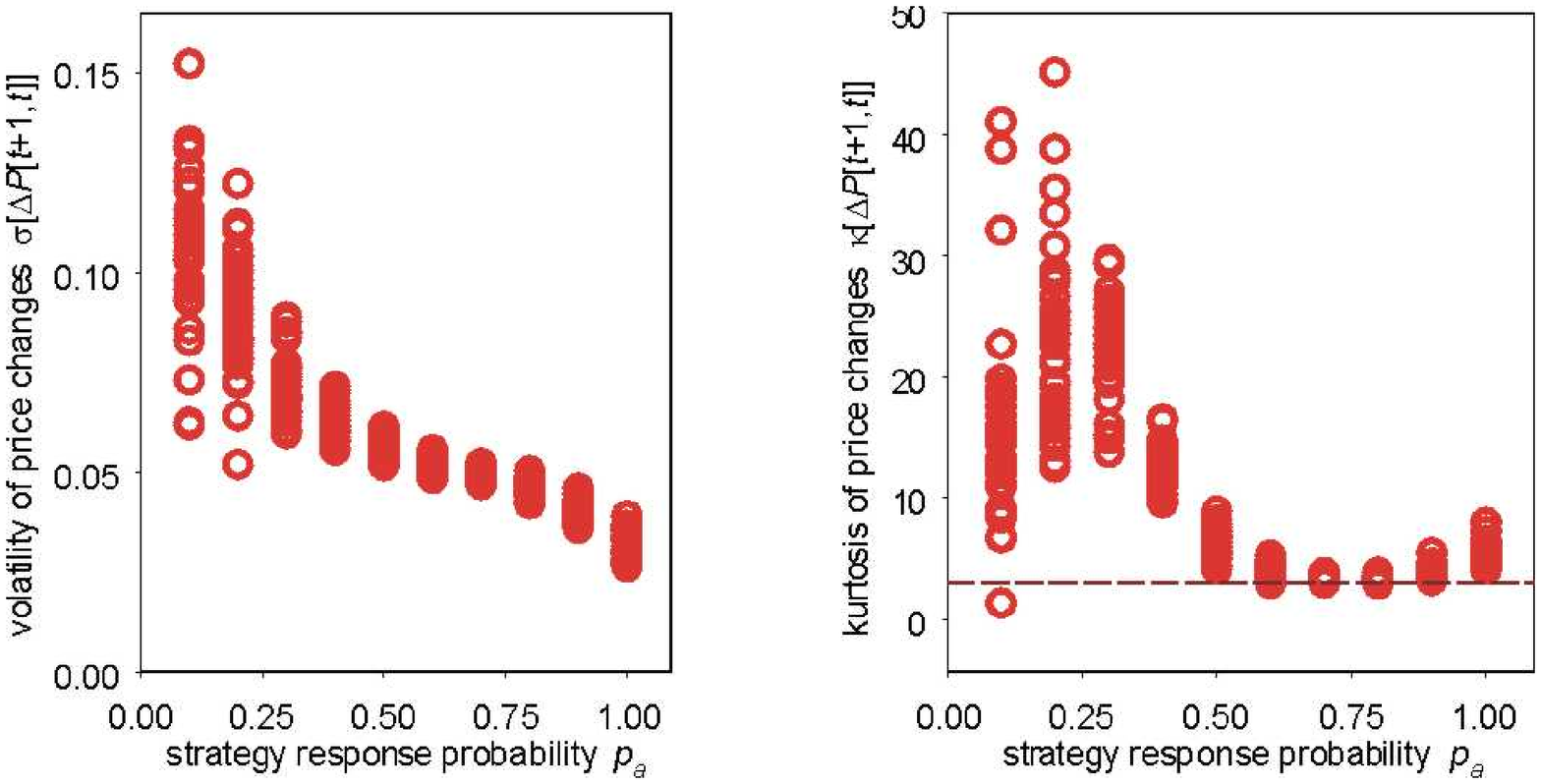}%
\caption{The volatility and kurtosis for a market model with incomplete
strategies, as a function of the probability $p_{a}$ that a strategy will have
a non-null action for a given global information state. Circles correspond to
values taken over the last 1000 timesteps of a 2000 timestep run. The model
parameters are $N=501$, $s=2$, $T=100$, and $r=4$, with binary ($\chi
=\operatorname*{sgn}$) strategy rewards.}%
\label{fig ternGW momes}%
\end{center}
\end{figure}

It is interesting to compare the statistical features of these models
featuring incomplete strategies, with the features of the GCMG. Figure
\ref{fig ternGW momes} shows the volatility and kurtosis of the price
movements $\Delta P\left[  t+1,t\right]  $ for a model with incomplete
strategies, as a function of the probability $p_{a}$ that a strategy has a
non-null action for a given global information state. The market model used
has the structure of the GCMG (N.B. for $p_{a}=1$, it \emph{is} the GCMG) and
features the behavior wherein agents maintain their previous course of action
when they encounter a null strategy action. As a consequence, the volatility
of the price movements rises as $p_{a}$ is reduced since it leads to more
persistence of agent action and consequently bigger asset price movements.
This affect can also be seen in the kurtosis which grows rapidly away from the
Gaussian case, since the probability of large moves increases with falling
$p_{a}$. [For more details on the nature of large movements in the GCMG, we
refer to Ref. \cite{uscrash}.]%

\section{Market clearing\label{sec Mkt making}}

In the preceding sections we have discussed mechanisms through which financial
market agents come to make trading decisions about whether to buy or sell an
asset. In Section \ref{sec price process} we made some suggestions about how
the combined action of all these agents' decisions produced a demand, and how
that demand then affected the movement of the asset price. So far we have not
explicitly mentioned a mechanism by which the market clears, i.e. by which
sellers of the asset meet buyers. The discussion of specific clearing
mechanisms amounts to a discussion of specific market microstructure. However
our goal here is to seek to be as general as possible, and hence not limit the
scope of our market models to any particular sector or style of financial
market. We thus limit ourselves here to a simplistic discussion of market-making.

The first important aspect of the models we have considered so far is that the
agents make a decision about whether to buy or sell assets at time $t$ based
only on the information up to (but not including) time $t$. The orders as such
are considered as `market orders' because they are unconditional on the actual
traded price $p\left[  t\right]  $. Orders whose execution depends on the
traded price $p\left[  t\right]  $ are known as `limit orders' and are beyond
the scope of the model types presented here.

The presence of only market orders in our model brings us to the second
important consideration: the number of buy orders in general does not balance
the number of sell orders, i.e. in general $D\left[  t\right]  \neq0$ since
all orders seek execution at time $t$ irrespective of the price. There are two
options then of how the market could function. The first option is to assume
that the number of assets sold by the agents must equal the number bought by
the agents. In this case the total number of assets traded is given by,
\[
v_{\text{assets}}\left[  t\right]  =\min\left[  n_{\text{buy}}\left[
t^{-}\right]  ,n_{\text{sell}}\left[  t^{-}\right]  \right]
\]
where $n_{\text{buy}}$ is the number of agents who seek to buy, and
$n_{\text{sell}}$ the number who seek to sell. With this system, the agents in
the majority group (buyers or sellers) only get their order partially
executed: for example if $n_{\text{buy}}=40$ and $n_{\text{sell}}=50$, the
sellers can each only sell $v_{\text{assets}}\left[  t\right]  /n_{\text{sell}%
}=40/50=80\%$ of a quanta of assets. This is the system proposed in Ref.
\cite{6bouchaudlong}. An alternative system assumes that there is a third
party, a market-maker, who operates between buyers and sellers and makes sure
that each and every order is fulfilled by taking a position in the asset
herself. We then have that:
\[
v_{\text{assets}}\left[  t\right]  =\max\left[  n_{\text{buy}}\left[
t^{-}\right]  ,n_{\text{sell}}\left[  t^{-}\right]  \right]
\]
and that the market-maker's inventory is:
\begin{equation}
\phi_{M}\left[  t\right]  =\phi_{M}\left[  t-1\right]  -D\left[  t^{-}\right]
\label{eq MM inventory}%
\end{equation}
The market-maker, unlike the agents who just seek to maximize their wealth,
has the joint goals of wealth maximization and staying market-neutral, i.e. as
close to $\phi_{M}\left[  t\right]  =0$ as possible. If the nature of the
market is highly mean-reverting, the second of these conditions will be
automatically true since $\sum_{j=0}^{t}D\left[  t\right]  \sim0$. In these
circumstances, the market-maker should employ an arbitrage-free price
formation process such as Equation \ref{eq price formation mult} or Equation
\ref{eq price formation add}. If on the other hand the market is not
mean-reverting, then the market-maker's position is unbounded. In these
circumstances a market-maker should employ a price-formation process such as
Equation \ref{eq price formation mult MM} which encourages mean reversion
through manipulation of the asset price based on $\sum_{j=0}^{t}D\left[
t\right]  $, as described in Section \ref{sec price process}. A market-maker
using a price formation such as Equation \ref{eq price formation mult MM} can
however be arbitraged by the agents. Whether the market-maker is actually
arbitraged is then purely dependent on the underlying dynamics of the market.

Let us demonstrate this by following Ref. \cite{5dollar} using a \$G\ref{eq
strat payoff posNWealth} model. For this model we have:
\begin{equation}
D\left[  t^{-}\right]  =A\left[  t-1\right]  -A\left[  t-2\right]
\label{eq demand dolGa}%
\end{equation}
as shown in Section \ref{sec payoff mods}. From Equation \ref{eq MM inventory}
we then get that $\phi_{M}\left[  t\right]  =\sum_{j=0}^{t-1}D\left[
t\right]  =\phi_{M}\left[  0\right]  -A\left[  t-1\right]  $ This immediately
tells us that the market-maker's inventory is bounded. The natural choice of
price formation process to use would then be the arbitrage-free case of, for
example, Equation \ref{eq price formation mult}. If we combine Equation
\ref{eq price formation mult} with Equation \ref{eq demand dolGa} we arrive
at:
\begin{equation}
p\left[  t\right]  =p\left[  0\right]  e^{\frac{A\left[  t-1\right]  }%
{\lambda}}\label{eq price dolG noMM}%
\end{equation}
We can thus see how the \$G\ref{eq strat payoff posNWealth} model produced the
tightly bounded, jumpy price series seen in Figure \ref{fig MM price vol}: the
price at time $t$ is no longer directly dependent on the price at time $t-1$
and is bounded by $p\left[  0\right]  e^{-\frac{N}{\lambda}}\leq p\left[
t\right]  \leq p\left[  0\right]  e^{\frac{N}{\lambda}}$.

Given the behavior of the \$G\ref{eq strat payoff posNWealth} model, the use
of the price formation process of Equation \ref{eq price formation mult MM}
seems rather unnecessary: the market maker does not need to manipulate the
price in order to stay market-neutral on average. However, let us demonstrate
the effect of using Equation \ref{eq price formation mult MM} as in Ref.
\cite{5dollar}. Combining Equations \ref{eq price formation mult MM} and
\ref{eq demand dolGa} with the market maker's inventory of $\phi_{M}\left[
t\right]  =-A\left[  t-1\right]  $ we arrive at:
\begin{equation}
p\left[  t\right]  =p\left[  t-1\right]  \exp\left[  \frac{1}{\lambda}A\left[
t-1\right]  +\left(  \frac{1}{\lambda_{M}}-\frac{1}{\lambda}\right)  A\left[
t-2\right]  \right] \label{eq price dolG MM}%
\end{equation}
Unlike Equation \ref{eq price dolG noMM}, Equation \ref{eq price dolG MM}
gives us again a multiplicative price process for finite values of
$\lambda_{M}$. Although it seems an inappropriate step, we would thus expect
that introducing Equation \ref{eq price formation mult MM} as a price
formation process would help regain some of the `stylized features' of a
financial market inherent from the multiplicative price process. However,
closer consideration of Equation \ref{eq price dolG MM} reveals something
different. Let us consider the case of $A\left[  t-2\right]  =A\left[
t-1\right]  $. In this case the price change is simply an increasing function
of $A\left[  t-1\right]  $. If more agents are long rather than short at time
$t$ ($A\left[  t-1\right]  >0$), then the price movement $\Delta P\left[
t,t-1\right]  $ is positive. The agents will then have benefitted from a
positive increment in their notional wealth from simply holding their
position. This manipulation of the (naive) market-maker leads to an unstable
state whereby the agents arbitrage the market-maker simply by holding their
long/short positions and watching the price rise/fall. This unstable state of
the market will occur with increasing probability as the agents become more
coordinated in their actions, i.e. as $m$ is decreased. For high $m$, where
there is no longer significant agent coordination and no crowding affects, the
\$G\ref{eq strat payoff posNWealth} model with Equation \ref{eq price dolG MM}
for the price formation is able to avoid the unstable attractor and thus
produce more `market-like' time-series as demonstrated in Ref. \cite{5dollar}.
We can also examine the changing wealth of the market-maker by simply
employing Equation \ref{eq agent wealth chng}. We know that $\phi_{M}\left[
t\right]  =-A\left[  t-1\right] $, so Equation \ref{eq agent wealth chng}
combined with Equation \ref{eq price dolG MM} gives us:
\begin{align}
\Delta W_{M}\left[  t+1,t\right]   &  =\phi_{M}\left[  t\right]  \Delta
P\left[  t+1,1\right] \nonumber\\
&  =-\frac{1}{\lambda}A\left[  t\right]  A\left[  t-1\right]  -\left(
\frac{1}{\lambda_{M}}-\frac{1}{\lambda}\right)  A\left[  t-1\right]
^{2}\label{eq MM wealth chng}%
\end{align}
In the case of $\lambda_{M}=\lambda$ as in Ref. \cite{5dollar}, we see can see
- by taking a time average of Equation \ref{eq MM wealth chng} - that whether
the market-maker's wealth increases or decreases with time will depend on the
dynamics of the market. If the market is persistent as in the unstable case at
low $m$, the market maker loses wealth since $\left\langle A\left[  t\right]
A\left[  t-1\right]  \right\rangle >0$. Conversely if the market is
anti-persistent, the market-maker will gain wealth since $\left\langle
A\left[  t\right]  A\left[  t-1\right]  \right\rangle <0$.

At first sight, it therefore seems that the \$G\ref{eq strat payoff
posNWealth} model can generate a `reasonably market-like' state wherein the
price is not unstable, it exhibits market-like dynamics and has a market-maker
who is able to make a profit from her role. However, in order to generate this
state it was necessary to use what seems to be an inappropriate market-making
mechanism and also push the model into the uncrowded phase. Neither of these
choices seem desirable in the quest for producing a realistic market model.

\section{Conclusion\label{sec Conclusion}}

Following the introduction to the community of the Minority Game, there have
been many proposed agent based market models, many of which exhibit the
`stylized facts' of financial market timeseries in certain parameter ranges.
It seems however that many of these market models include one or more basic
assumptions which seem implausible when compared to the actions of real
financial market agents.

With the first `wave' of market models it seemed remarkable that so many of
the stylized features of real financial timeseries were reproduced so well.
With more and more market model contenders entering the community, each with a
quite diverse set of assumptions, it started to look as though it was easy to
find models which would reproduce realistic market-like dynamics. It has been
proposed \cite{9bouchaudprague} that general systems which have activity based
on a waiting time problem, will all enjoy the scaling properties seen in
financial market data. Notwithstanding this observation, we believe that it is
not at all obvious how one can produce a self-consistent market model which
embodies, in a \emph{non-stochastic} way, the behavior of realistic financial
market agents, and yet can reproduce the `stylized facts' of real financial
markets. As we have shown in this work, adaptation of the MG payoff structure
to incorporate a more realistic objective function results in a model
increasingly dominated by trending as agents' resources are increased. As a
result of this we are forced to look to a wider class of agent behavior and
heterogeneity in order to recover the dynamical behavior we are looking for.

This paper has provided a discussion of different aspects of market models and
how they interact with each other. Instead of proposing a unique new market
model of our own (thereby adding to the increasing family in the literature),
we have examined the different features one may want to include in a market
model, and then discussed what their effect would be. We hope that the
resulting discussion has managed to provide a basic of toolbox of model
features, brought together by a common formalism. Perhaps some optimal
combination of a subset of these model features will provide a new
self-consistent, realistic yet minimal market model. However it is likely that
one needs a diversity not only of agent parameter values but also of agent
behavior, if one is intent on forming \emph{the} market model.

Finally we emphasize that this paper falls into the category of an
\emph{interim progress report} on our research in this area to date - a
synthesis of our work over the past few years and thoughts, particularly in
light of recent proposals for modified agent-based models. We would therefore
welcome any comments and, in particular, criticisms of our results and
opinions. We also thank the many people with whom we have enjoyed discussions
about financial modelling, both within and outside the Econophysics community.


\newpage

\begin{thebibliography}{99}
\bibitem{1arthur}W.B. Arthur, Science \textbf{284}, 107 (1999). See also
references therein.

\bibitem{2econophysics}See econophysics website www.unifr.ch/econophysics. See
also J. Bouchaud and M. Potters, \emph{Theory of Financial Risks} (Cambridge
University Press, Cambridge, 2000); R. Mantegna and H. Stanley,
\emph{Econophysics} (Cambridge University Press, Cambridge, 2000); B.
Huberman, P. Pirolli, J. Pitkow, and R. Lukose, Science \textbf{280}, 95
(1998); T. Lux and M. Marchesi, Nature \textbf{397}, 498 (1999); R.G. Palmer,
W.B. Arthur, J.H. Holland, B. LeBaron and P. Tayler, Physica D \textbf{75},
264 (1994).

\bibitem{3challet}D. Challet and Y.C. Zhang, Physica A \textbf{246}, 407
(1997); Physica A \textbf{256} 514 (1998).

\bibitem{4liege}P. Jefferies, N.F. Johnson, M. Hart, and P.M. Hui, Eur. Phys.
J. B \textbf{20}, 493 (2001). This paper was presented at the APFA2 Conference
in Liege, 2000.

\bibitem{5dollar}J.V. Andersen and D. Sornette, cond-mat/0205423.

\bibitem{6bouchaudlong}I. Giardiana and J.P. Bouchaud, cond-mat/0206222.

\bibitem{7occf}Sets of past and present working papers can be viewed on the
OCCF website www.occf.ox.ac.uk.

\bibitem{challet}D. Challet, M. Marsili and Y.C. Zhang, Physica A
\textbf{299}, 228 (2001).

\bibitem{15marsili}M. Marsili, Physica A \textbf{299}, 93 (2001).

\bibitem{farmer}J.D. Farmer and S. Joshi, cond-mat/0012419.

\bibitem{8dublin}N.F. Johnson, M. Hart, P.M. Hui, and D. Zheng, Int. J. Theo.
Appl. Fin. \textbf{3}, 443 (2000). This paper was presented at the APFA1
Conference in Dublin, 1999.

\bibitem{usvarL1}N.F. Johnson, P.M. Hui, D. Zheng, and C.W. Tai, Physica A
\textbf{269}, 493 (1999).

\bibitem{usvarL2}P. Jefferies, D. Lamper and N.F. Johnson, submitted to
Physica A (2002). This paper is based on work presented at the APFA3
Conference in London, 2001.

\bibitem{uspre}P. Jefferies, M. Hart, and N.F. Johnson, Phys. Rev. E
\textbf{65}, 016105 (2002).

\bibitem{9bouchaudprague}I. Giardina, J.P. Bouchaud and M. Mezard, Physica A
\textbf{299}, 28 (2001).

\bibitem{10contbf}J.P. Bouchaud and R. Cont, Eur. Phys. J. B \textbf{6}, 543
(1998); J.D. Farmer, adap-org/9812005.

\bibitem{11empiricaldollar}T. Chordia, R. Roll and A. Subrahmanyam, in press
J. Fin. Econ. (2001); J. Lakonishok, A. Shleifer, R. Thaler and R. Vishny, J.
Fin. Econ. \textbf{32}, 23 (1991); S. Maslov and M. Mills, Physica A
\textbf{299}, 234 (2001); D. Challet and R. Stinchcombe, Physica A
\textbf{300}, 285 (2001).

\bibitem{12crowd}M. Hart, P. Jefferies, N.F. Johnson and P. M. Hui, Physica A
\textbf{298}, 537 (2001); N.F. Johnson, M. Hart, and P.M. Hui, Physica A
\textbf{269}, 1 (1999); M. Hart, P. Jefferies, N.F. Johnson, and P. M. Hui,
Phys. Rev. E \textbf{63}, 017102 (2000); N.F. Johnson, P.M. Hui, D. Zheng, and
M. Hart, J. Phys. A: Math. Gen. \textbf{32}, L427 (1999); T.S. Lo, S.W. Lim,
P.M. Hui and N.F. Johnson, Physica A \textbf{287}, 313 (2000); P. Jefferies,
M. Hart, N.F. Johnson, and P.M. Hui, J. Phys. A \textbf{33}, L409 (2000); M.
Hart, P. Jefferies, P.M. Hui and N.F. Johnson, Eur. Phys. J. B \textbf{20},
547 (2000).

\bibitem{13marsili}M. Marsili and D. Challet, cond-mat/0004376.

\bibitem{14genetic}N.F. Johnson, P.M. Hui, R. Jonson and T.S. Lo, Phys. Rev.
Lett. \textbf{82}, 3360 (1999); T.S. Lo, P.M. Hui, and N.F. Johnson, Phys.
Rev. E \textbf{62}, 4393 (2000); P.M. Hui, T.S. Lo, and N.F. Johnson, Physica
A \textbf{288}, 451 (2000); N.F. Johnson, D.J.T. Leonard, P.M. Hui, and T.S.
Lo, Physica A \textbf{283}, 568 (2000).

\bibitem{uscrash}D. Lamper, S.D. Howison and N.F. Johnson, Phys. Rev. Lett.
\textbf{88}, 017902 (2002).
\end{thebibliography}
\end{document}